\documentclass{aa}  
\usepackage{natbib}
\bibpunct{(}{)}{;}{a}{}{,} 
\usepackage{graphicx}
\usepackage{txfonts}
\usepackage{multirow}
\usepackage{enumerate}
\begin{document}

   \title{A new insight into the Galactic potential: A simple secular model for the evolution of binary systems in the solar neighbourhood}

   \author{J. A. Correa-Otto
          \inst{1,2}
          \and
          M. F. Calandra\inst{1}
          \and
          R. A. Gil-Hutton\inst{1,2}
          }
   \institute
             {Grupo de Ciencias Planetarias, Complejo Astron\'omico El Leoncito, UNLP, UNC, UNSJ, CONICET, Av. España 1512 sur, J5402DSP San Juan, Argentina\\
             \email{jcorrea@casleo.gov.ar} \\
         \and
            Universidad Nacional de San Juan, J. I. de la Roza 590 oeste, 5400 Rivadavia, San Juan, Argentina.\\  }

   \date{Received ; accepted }

 
  \abstract
   {Among the main effects that the Milky Way exerts in binary systems, the Galactic tide is the only one that is not probabilistic and can be deduced from a potential. Therefore, it is possible to perform an analysis of the global structure of the phase space of binary systems in the solar neighbourhood using the Galactic potential.}
   {The aim of this work is to obtain a simple model to study the collisionless dynamical evolution of generic wide binaries systems  in the solar neighbourhood.}
   {Through an  averaging process, we reduced the three-dimensional potential of the Galaxy to a secular one-degree of freedom model. The accuracy of this model was tested by comparing its predictions with  numerical simulations of the exact equations of motion of a two-body problem disturbed by the Galaxy.}
   {Using the one-degree of freedom model, we developed a detailed dynamical study, finding that the secular Galactic tide period changes as a function of the separation of the pair, which also gives a  dynamical explanation for the arbitrary classification between "wide" and "tight" binaries. Moreover, the secular phase space for a generic gravitationally bound pair is similar to the dynamical structure of a Lidov-Kozai resonance, but surprisingly this structure is independent of the masses and semimajor axis of the binary system. Thus, the Galactic potential is able to excite the initially circular orbit of binary systems to high values of eccentricity, which has important implications for studies of binary star systems (with and without exoplanets), comets, and Oort cloud objects.}
   {}

   \titlerunning{A model for the evolution of binary systems in the solar neighbourhood}
   \maketitle

%


\section{Introduction}\label{intro}

The binary systems in the solar neighbourhood are commonly associated with gravitationally bounded stars. However, the Sun and a comet or an Oort cloud object can also be considered a binary system in the framework of a restricted two-body problem. For any of these pairs, their orbits mainly change by the influence of Galactic tides and encounters with close passing stars and molecular clouds (Brunini 1995; Eggers \& Woolfson 1996; Levison \& Dones 2001; Fouchard et al. 2006; Jiang \& Tremaine 2010; Kaib et al. 2011, 2013).

The effect exerted by the Milky Way on a binary system depends on its orbit, since its semimajor axis ($a$) regulates the influence of the Galactic environment. Binary systems with a large separation are weakly bound by self-gravity, and therefore they are the most disturbed by the Galaxy (Heggie 1975; Bahcall et al. 1985; Jiang \& Tremaine 2010). These configurations with $a >$ 1000 au are called "wide binaries" (Roell et al. 2012) and a system with smaller semimajor axis is considered a "tight or close" binary. This limit ($\sim$ 1000 au) is defined empirically, but there is not an analytic deduction that supports it.

In the solar neighbourhood, our Galaxy disturbs a binary system with two main effects: first, the tidal field of the Milky Way and, second, the gravitational perturbation from passing stars or other perturbers (e.g. molecular clouds). However, while the Galactic tidal field is derived from an analytical potential, the effects of encounters caused by passing stars (or other object) correspond to a stochastic perturbation on the pair. 

Therefore, because of the random influence of the Milky Way in wide pairs, almost all the works carried out in this area are statistical studies. However, despite the dynamic importance of a theoretical analysis about the effect of the Galactic tide on a binary system, it is difficult to find these types of studies in the literature.

Heisler \& Tremaine (1986) performed an analytical study of the Galactic potential for the restricted two-body problem (i.e. Sun-comet), but they focused on the lost rate of minor bodies due to the Galactic tide and stellar encounters. We have not found a specific study about the dynamic portrait of the secular phase space of a binary system disturbed by the gravitational potential of the Milky Way.

The importance of a detailed dynamical study about these systems relies on the possibility of simplifying the Hamiltonian.  Such a mathematical procedure allows us to reduce the equations of motion, with the consequent low computational cost, despite the diversity of systems and initial configurations that can be analysed.

The aims of the present work are twofold: first, to obtain a simple model for the secular evolution of a pair of objects that are gravitationally bound and located in the solar neighbourhood, and  second,  through that model, to perform a detailed study of the effects of the Galactic gravitational potential in generic binary systems (i.e. star-star, Sun-comet, etc.).

In Section \ref{model} we deduce the model that has one-degree of freedom. In section \ref{num} the model is used to find the position of the equilibrium points of the Hamiltonian. In section  \ref{fase} a detailed analysis of the phase space of a binary system is presented. An application of the model is in Section \ref{referi}. Conclusions close the paper in Sect. \ref{conclu}.

\section{The model for the Galactic potential}\label{model}

We allow $m_1$ and $m_2$ to be the masses of the components of a binary system (i.e. star-star, star-planet, Sun-comet, etc.), which are negligible compared with the mass of the Galaxy. The motion of such a binary system in the Galaxy can be described  by the use of the Hill approximation (Heggie 2001; Binney \& Tremaine 2008). We consider a non-inertial Cartesian astrocentric coordinate system $(x, y, z)$ with origin in the main body ($m_1$). This system is at a distance $R_g$ from the Galactic centre, and corotates with the Galaxy. The $z-axis$ is perpendicular to the Galactic plane and points towards the south Galactic pole, the $y-axis$ points in the direction of Galactic rotation, and the $x-axis$  points radially outwards from the Galactic centre.

Then, assuming a symmetric potential on the plane $z = 0$, the Galactic tide disturbs the binary system according to the Hamiltonian,

\begin{equation}
\begin{array}{ccl}
\mathcal{K} &=& \dfrac{1}{2} (\dot{x}^2+\dot{y}^2+\dot{z}^2) - \dfrac{\mu}{\sqrt{x^2+y^2+z^2}}  \\ 
            &&  \\
            &-&  2 \Omega_G (x \dot{y} - y \dot{x}) - 2 \Omega_G A_G x^2 + \dfrac{\nu_G^2 z^2 }{2}  \,  \rm{,} \label{eq1} \\
\end{array}
\end{equation}

\noindent where $\mu= \mathcal{G}  (m_2 + m_1)$, $\mathcal{G}$ is the gravity constant and $x$, $y$, $z$, $\dot{x}$, $\dot{y}$, $\dot{z}$ are the astrocentric components of position and velocity of the secondary body $m_2$ around $m_1$. Moreover, $\Omega_G$, $A_G$, and $\nu_G$ are the angular speed of the Galaxy, Oort constant, and  frequency for small oscillations in $z$, respectively. For $R_g = 8 $ kpc (i.e. approximately the distance from the Galactic centre to the Sun) their values are Jiang \& Tremaine (2010)

\begin{equation}
\begin{array}{ccl}
\Omega_G &\cong& 3.017 \times 10^{-8} \, yr^{-1}\,  \rm{,} \\ 
A_G &\cong& 1.513 \times 10^{-8} \, yr^{-1}\,  \rm{,} \\ 
\nu_G &\cong& 7.258 \times 10^{-8} \, yr^{-1} \,  \rm{.} \label{eq2}
\end{array}
\end{equation}

\noindent  The equations of motion for this Hamiltonian are (see Jiang \& Tremaine (2010))

\begin{equation}
\begin{array}{cclccl}
\dot{x} &=& \dfrac{\partial \mathcal{K}}{\partial \dot{x}} , & \ddot{x} &=& -\dfrac{\partial \mathcal{K}}{\partial x} \,  \rm{,}\\ 
   \\
\dot{y} &=& \dfrac{\partial \mathcal{K}}{\partial \dot{y}} , & \ddot{y} &=& -\dfrac{\partial \mathcal{K}}{\partial y}  \,  \rm{,} \\ 
   \\
\dot{z} &=& \dfrac{\partial \mathcal{K}}{\partial \dot{z}} , & \ddot{z} &=& -\dfrac{\partial \mathcal{K}}{\partial z} \,  \rm{.} \label{eq3}
\end{array}
\end{equation}

In dynamical studies it is convenient to work with orbital elements instead of Cartesian coordinates. Once again, we place the centre of the coordinate system in $m_1$, and then the orbital elements correspond to the astrocentric orbit of $m_2$ around $m_1$. We define the orbital elements as follows: the size and form of the orbit are defined by the semimajor axis $a$ and the eccentricity $e$, respectively. The mean anomaly $M$ indicates the position of $m_2$ in the orbit. The reference plane is the mid-plane of the Galaxy, and its intersection with the orbital plane of $m_2$ defines the position of the ascending node. The direction radially outwards from the Galactic centre (i.e. $x-axis$) is taken as reference to define the angular position of the ascending node $\Omega$, where the angle is measured in anti-clockwise sense (i.e. towards the positive $y-axis$). The inclination $I$ is defined with respect to the reference plane, and the argument of pericentre $\omega$ is measured from the ascending node in an anti-clockwise sense. For these astrocentric orbital elements, we can define the Hamiltonian as


\begin{equation}
\mathcal{K} = -\dfrac{\mu}{2 a} - 2 \Omega_G \sqrt{\mu a (1 - e^2)} \cos I - 2 \Omega_G A_G x^2 + \dfrac{\nu_G^2 z^2 }{2}  \,  \rm{,} \label{eq4}
\end{equation}

\noindent where 

\begin{equation}
\begin{array}{ccl}
x &=& a \lbrace \cos \Omega [  (\cos E  -  e) \cos \omega -  \sqrt{1-e^2} \sin E \sin \omega] \\
  &-& \cos I \sin \Omega [  (\cos E  -  e) \sin \omega +  \sqrt{1-e^2} \sin E \cos \omega ] \rbrace \,  \rm{,} \\
     \\
z &=& a \sin I  [  (\cos E  -  e) \sin \omega + \sqrt{1-e^2} \sin E \cos \omega ] \,  \rm{,} \label{eq5}
\end{array}
\end{equation}

\noindent and $E$ is the eccentric anomaly, which is an implicit function of the mean anomaly  ($M=E- e \sin E$).

The first term in equation (\ref{eq4}) corresponds to the two-body contribution, $\mathcal{K}_0$. The second term in eq. (\ref{eq4}) is the $z$ component  of the angular momentum multiplied by the angular speed of the Galaxy ($\Omega_G$) and it is included because we are working in a non-inertial frame. The influence of this term ($\mathcal{K}_1$) is small compared with the two-body term $\mathcal{K}_0$. Finally, the last two terms correspond to the tidal perturbation of the Galaxy, which are still smaller than the previous terms, so both are considered a perturbation called $\mathcal{K}_2$. Then, we can schematically write the Hamiltonian as a disturbed two-body problem,

\begin{equation}
\mathcal{K} = \mathcal{K}_0 +\mathcal{K}_1+\mathcal{K}_2 \,  \rm{.} \label{eq6}
\end{equation}

Considering the  distant-tide approximation Jiang \& Tremaine (2010), it is possible to write the effects of the Galactic tidal field (i.e. $\mathcal{K}_2$) about $R_g=$ 8 kpc using a Taylor series expansion. We truncate the series expansion at second order in $r = \sqrt{x^2+y^2+z^2}$, as in many other works  (Heisler \& Tremaine 1986; Jiang \& Tremaine 2010; Kaib et al. 2013). Although it is possible to consider higher orders of the series, their effects are negligible for the range of distance ($r$) considered in this paper (Sect. \ref{scope}) and also for larger separations between the pair Jiang \& Tremaine (2010). 

However, the orbital elements are not canonical variables, which makes the application of the canonical perturbation theory  difficult (Morbidelli 2002; Ferraz-Mello 2007), despite the complex equations of variation of the orbital elements. Then, a better choice is to work with the canonical action-angles variables of Delaunay associated with the orbital elements defined above written as

\begin{equation}
\begin{array}{ccll}
L &=& \sqrt{\mu a} \, , & M  \,  \rm{,} \\ 
   \\
G &=& L \sqrt{1 - e^2} \, , & \omega \,  \rm{,}  \\ 
   \\
H &=& G \cos I \, , & \Omega \,  \rm{.} \label{eq7}
\end{array}
\end{equation}

\noindent In these variables the Hamiltonian is defined as

\begin{equation}
\begin{array}{ccl}
\mathcal{K}_0 &=& - \frac{1}{2}\mu^2 L^{-2}  \,  \rm{,}    \\ 
   \\
\mathcal{K}_1 &=& - 2 \Omega_G H  \,  \rm{,}  \\ 
   \\
\mathcal{K}_2 &=& - 2 \Omega_G A_G x^2 + \dfrac{1}{2} \nu_G^2 z^2  \,  \rm{,} \label{eq8}
\end{array}
\end{equation}

\noindent where

\begin{equation}
\begin{array}{ccl}

x & = & \dfrac{L^2}{\mu} \bigg[  \cos E \left(  \cos \omega - \dfrac{H}{G} \sin \omega   \sin \Omega \right)  \\ 
&& \\
  & - & \dfrac{G}{L} \sin E \left( \sin \omega \cos \Omega + \dfrac{H}{G} \cos \omega  \sin \Omega \right)  \\
  && \\
  & + &  \sqrt{1- \left( \dfrac{G}{L}  \right)^2} \left(\dfrac{H}{G} \sin \omega  \sin \Omega - \cos \omega \cos \Omega \right)  \bigg] \,  \rm{,} \\
\\

     \\
z & = & \dfrac{L^2}{\mu} \sqrt{1- \left( \dfrac{H}{G}  \right)^2}  \bigg[   \cos E \sin \omega \\
 & & \\
  & +&  \dfrac{G}{L} \sin E \cos \omega  -  \sqrt{1- \left( \dfrac{G}{L}  \right)^2} \sin \omega \bigg]  \,  \rm{,} \label{eq9}
\end{array}
\end{equation}

\noindent  and with these variables, the equations of variation of the orbital elements are the following Delaunay equations: 

\begin{equation}
\begin{array}{cclccl}
\dot{M} &=& \dfrac{\partial \mathcal{K}}{\partial L} \, , & \dot{L} &=& -\dfrac{\partial \mathcal{K}}{\partial M} \,  \rm{,} \\ 
   \\
\dot{\omega} &=& \dfrac{\partial \mathcal{K}}{\partial G} \, , & \dot{G} &=& -\dfrac{\partial \mathcal{K}}{\partial \omega} \,  \rm{,}  \\ 
   \\
\dot{\Omega} &=& \dfrac{\partial \mathcal{K}}{\partial H} \, , & \dot{H} &=& -\dfrac{\partial \mathcal{K}}{\partial \Omega} \,  \rm{.} \label{eq10}
\end{array}
\end{equation}

For the objectives of this paper, it is convenient to write the eq. (\ref{eq9}) as
\begin{equation}
\begin{array}{ccl}

x & = & a x_1 \, \\ 
\\
z & = & a z_1  \,  \rm{,} \label{eq9b}
\end{array}
\end{equation}

\noindent where $x_1$ and $z_1$ are functions of the orbital elements $e$, $I$, $M$, $\omega,$ and $\Omega$ with their range of values between -1 and 1. Moreover, the derivative of these functions with respect to the action variables $L$, $G,$ and $H$ can also be written in a simplified form as

\begin{equation}
\begin{array}{ccl}

\dfrac{\partial x}{\partial \chi} & = & \sqrt{\dfrac{a}{\mu}} x_\chi \, \\ 
\\
\dfrac{\partial z}{\partial \chi} & = & \sqrt{\dfrac{a}{\mu}} z_\chi  \,  \rm{,} \label{eq9c}
\end{array}
\end{equation}

\noindent with $\chi$ representing any of the action variables. The functions $x_\chi$ and $z_\chi$ are dependent of the orbital elements $e$, $I$, $M$, $\omega,$ and $\Omega$, and they are bounded in the range $\in$ (-1, 1).

Finally, the limit for the disturbed two-body problem  depends on the semimajor axis, since for small values of $a$ the perturbations of the Galaxy ($\mathcal{K}_1$ and $\mathcal{K}_2$) are too small. However, with the increase of the semimajor axis such approximation is no longer valid because of the increase of the disturbing terms (i.e. we leave the regime of a disturbed two-body problem).

\subsection{The one-degree of freedom model} \label{The one-dimensional approximation model} 

The Hamiltonian (\ref{eq6}) can be written as

\begin{equation}
\mathcal{K} = \mathcal{K}_{g0} + \epsilon \, \mathcal{K}_{g1}   \,  \rm{,} \label{eq6b}
\end{equation}

\noindent with,

\begin{equation}
\begin{array}{ccl}
\mathcal{K}_{g0} &=& - \dfrac{\mu^2}{2 L^{2}}  - 2 \Omega_G H  \,  \rm{,}    \\ 
   \\
\epsilon &=& \left(  \dfrac{a}{r_J} \right) ^3  \,  \rm{,}  \\ 
   \\
\mathcal{K}_{g1} &=& - \dfrac{\mu}{2 \, a^3}  x^2 + \dfrac{\beta \, \mu}{4 \, a^3} z^2  \,  \rm{,} \label{eq6c}
\end{array}
\end{equation}

\noindent where

\begin{equation}
\beta = \dfrac{\nu^2}{2 \Omega_G A_G} = 5.77  \,  \rm{} \label{eq6d}
\end{equation}

\noindent is a constant and 

\begin{equation}
r_J = \left(  \dfrac{\mu }{4 \Omega_g A_g } \right) ^{1/3}  \,  \rm{} \label{eq6dx}
\end{equation}

\noindent is the radius of Jacobi of a binary system in the tidal field of the Galaxy Jiang \& Tremaine (2010). Considering a three-body system composed by the binary system and the Galaxy, the tidal radius $r_J$ is the position of the stationary Lagrange points L$_1$, which defines the Roche lobe of the binary. Here, $\mathcal{K}_{g0}$ is the integrable approximation, $\mathcal{K}_{g1}$ the disturbing function or perturbation. and $\epsilon$ a small parameter. Thus, we can represent the Hamiltonian (\ref{eq6}) as the Hamiltonian of a quasi-integrable system (Morbidelli 2002; Ferraz-Mello 2007). Moreover, the quasi-integrable Hamiltonian (\ref{eq6b}) can be written in the simplified form,

\begin{equation}
\mathcal{K}(L,H,G,M,\Omega,\omega) = \mathcal{K}_{g0} (L,H) + \epsilon \, \mathcal{K}_{g1}(L,H,G,M,\Omega,\omega)  \,  \rm{,} \label{eq6e}
\end{equation}

\noindent where 

\begin{equation}
\begin{array}{ccl}
f_o &=&  \dfrac{\partial \mathcal{K}_0}{\partial L} = n \,  \rm{,}    \\ 
   \\
f_n &=& \dfrac{\partial \mathcal{K}_0}{\partial H} = -2 \Omega_G  \,  \rm{}  \label{eq6f}
\end{array}
\end{equation}

\noindent are the frequencies of the  integrable approximation and $n$ the mean motion of the binary system. Therefore, we can perform a first-order averaging of the Hamiltonian over the fast angles $M$ and $\Omega$. This averaging is performed by the Hori method  at first order  (Morbidelli 2002; Ferraz-Mello 2007), which takes  advantage of the Lie series and is similar to the transformation of Von Zeipel-Brouwer (Brouwer \& Clemence 1961; Naoz et al.2013).

The averaged Hamiltonian can be written as

\begin{equation}
\mathcal{K^*}(G^*,\omega^*;L^*,H^*) = \mathcal{K^*}_{g0} (L^*,H^*) + \epsilon \, \mathcal{K^*}_{g1}(G^*,\omega^*;L^*,H^*)  \,  \rm{,} \label{eq6g}
\end{equation}

\noindent where $\mathcal{K^*}_{g0}$ has the same functional form that $\mathcal{K}_{g0}$, and $\mathcal{K^*}_{g1}$ define the homologic equation,

\begin{equation}
\mathcal{K^*}_{g1}= \mathcal{K}_{g1} + \lbrace \mathcal{K}_{g1} , \chi_g \rbrace \,   \rm{,}  \label{eq6h}
\end{equation}

\noindent with $\chi_g$ as the generating function. 

In order to solve the homologic equation we take advantage of the Hamiltonian (\ref{eq6}), which is periodic in the angles $M$ and $\Omega$,  the disturbing function can thereby be expanded in Fourier series,

\begin{equation}
\mathcal{K}_{g1}=  \sum_{\textbf{n} \in \textbf{Z}^3} c_\textbf{n} \, \rm{exp}^{i\textbf{n}.\textbf{q}}   \,  \rm{,}    \label{eq6i}
\end{equation}

\noindent where $\textbf{q}=(M,\Omega,\omega)$ and $\textbf{n}$ is an integer vector. Then, for a similar expansion of the homologic equation, the averaged disturbing function is simply $\mathcal{K^*}_{g1}=  c_0$. The corresponding coefficient $c_0$ of the Fourier expansion is

\begin{equation}
\mathcal{K^*}_{g1} = c_0  =  \dfrac{1}{4\pi^2} \int_0^{2\pi} \int_0^{2\pi} \mathcal{K}_{g1} \, dM \, d\Omega   \,  \rm{,} \label{eq6j}
\end{equation}

\noindent or using the implicit relation $M=E -e \, sin E$ and its differential form $dM=1-e \, \rm{cos} E \, dE,$ we can write

\begin{equation}
\mathcal{K^*}_{g1} = c_0  =  \dfrac{1}{4\pi^2} \int_0^{2\pi} \int_0^{2\pi} \mathcal{K}_{g1} \, (1-e \, \rm{cos} E) \, dE \, d\Omega  \,  \rm{.} \label{eq6k}
\end{equation}

\noindent For more details about the averaging process, see Chapter 2 of Morbidelli (2002).

Then, the averaging disturbing function is defined as

\begin{equation}
\begin{array}{ccl}

\mathcal{K}_{g1}^* & = & \dfrac{\mu^2}{4\,{L^*}^2} \bigg\lbrace  \dfrac{\alpha}{{G^*}^2}  [ (2{L^*}^2-{G^*}^2) ({G^*}^2-{H^*}^2) \\
 & & \\
  & - &  ({L^*}^2-{G^*}^2) ({G^*}^2-{H^*}^2)  \cos{2\omega^*} ] \\
   & & \\
  & - & (2{L^*}^2-{G^*}^2) \bigg\rbrace   \,  \rm{,}
\end{array}
 \label{eq61}
\end{equation}

\noindent with $\alpha = 0.5 (1 + \beta)$. In terms of astrocentric orbital elements, the mean-mean or averaged disturbed function can be expressed as

\begin{equation}
\mathcal{K}_{g1}^*  \, =  \, \dfrac{\mu}{4\,{a^*}}  [ \alpha (\sin{{I^*}})^2  (1+{e^*}^2 -   {e^*}^2 \cos{2\omega^*}) -  (1+{e^*}^2)  ]  \,  \rm{,}
 \label{eq61b}
\end{equation}

\noindent where $a^*$, $e^*$, $I^*$,  $\omega^*$ are the mean-mean orbital elements averaged in the fast angles $M$ and $\Omega$.

From the averaging process we obtain new canonical variables ($L^*$, $G^*$, $H^*$, $M^*$, $\omega^*$, $\Omega^*$) such that the transformed Hamiltonian is ${K^*}$=$\mathcal{K^*}$($G^*$, $\omega^*$; $L^*$, $H^*$) and $M^*$ and $\Omega^*$ are cyclic. The associated momenta $L^*$ and $H^*$ are then  new constants of motion, and the system is reduced to a model with one-degree of freedom for the canonical variables ($G^*$, $\omega^*$). Thus, the mean-mean Hamiltonian, which defines our one-degree of freedom model, is

\begin{equation}
\begin{array}{ccl}
\mathcal{K}^*  = \mathcal{K}_{g0}^* + \epsilon \mathcal{K}_{g1}^*  \,  \rm{,}
\end{array}
 \label{eq61c}
\end{equation}

\noindent with 

\begin{equation}
\begin{array}{ccl}
\mathcal{K}_{g0}^*  = - \dfrac{\mu^2}{2 (L^*)^{2}}  - 2 \Omega_G H^* \,  \rm{.}
\end{array}
 \label{eq61d}
\end{equation}

\noindent Owing to the conservation of $H^*$ ($H^*=L^* \sqrt{1-{e^*}^2} \cos {I^*}$), there is a coupling between the eccentricity and inclination, and thereby the dynamical behaviour of the system seems to be similar to the Lidov-Kozai resonance (Lidov 1961; Kozai 1962; Naoz 2016).

\subsubsection{Scope of the approximation}\label{scope}

The scope of our quasi-integrable Hamiltonian (\ref{eq6}) is defined by the weight of the disturbing function.  The second term of the right part in eq. (\ref{eq6}) must be a small perturbation over the integrable Hamiltonian $\mathcal{K}_{g0}$.

  \begin{figure}
\centering
\includegraphics [width=0.45\textwidth] {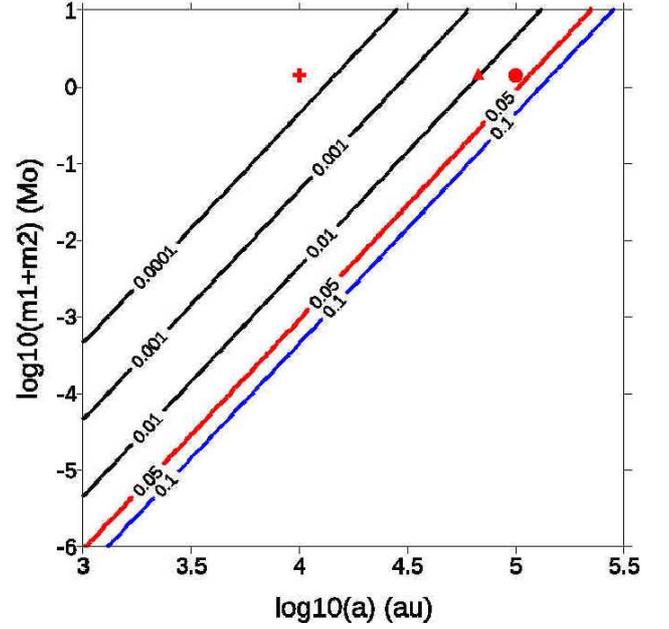} 
\caption{{\small Level curves of constant $\epsilon$ in the plane ($a$, $m_1+m_2$). The dotted red line indicates the limit for the integrable approximation. The approximation fails for the binary systems below the dashed blue line. The red symbols (cross, triangle, and circle) represent the position of three binary systems (see Sect. \ref{standard}). }}
\label{figx}
\end{figure}

The integrable Hamiltonian has two terms and working algebraically we obtain

\begin{equation}
\begin{array}{ccl}
\mathcal{K}_{g0}  = - \dfrac{\mu}{2 a} (1 + 4 \, J(e,i) \, \mu^{-0.5} \, \Omega_G \, a^{1.5}) \,  \rm{,}
\end{array}
 \label{eq71b}
\end{equation}

\noindent where $J$ is a function with values in the range $\in$ (-1,1). The functional form of  the second term in the parenthesis of the right side in eq. (\ref{eq71b}) can be approximately by its dependence on $a$ because the range of values of the parameter $\mu$ is smaller than the range of the semimajor axis. So, the second term in the parenthesis becomes comparable to 1 for $a \sim 10^5$ au. In this way we can approximate $\mathcal{K}_{g0}  \propto \mu a^{-1}$. On the other hand, the perturbation can be written using the simplified forms  (\ref{eq9b}) and (\ref{eq9c}), 

\begin{equation}
\epsilon \, \mathcal{K}_{g1} = \left(  \dfrac{a}{r_J} \right) ^3 \dfrac{\mu}{4\,a} (- 2  x_1^2 + 5.77 z_1^2 ) \,  \rm{,} \label{eq71c}
\end{equation}

\noindent and then, $\mathcal{K}_{g1} \propto \mu a^{-1}$. Therefore, the integrable Hamiltonian and the disturbing function are comparable terms, being $\epsilon = a^3 r_J^{-3}$  who defines the weight of the perturbation. The limit for the disturbing regime is difficult to define, but we can consider $\epsilon \sim $ 0.05  as the upper limit for the quasi-integrable approximation.

However, the small parameter only depends on the semimajor axes and the masses

\begin{equation}
\epsilon = \left(  \dfrac{a}{r_J} \right) ^3 = \dfrac{4 \Omega_G A_G}{\mathcal{G}} \dfrac{a^3}{m_1+m_2}  \,  \rm{.} \label{eq71x}
\end{equation}

\noindent Thus, the scope of the one-degree of freedom model is defined by two parameters.  Figure \ref{figx} shows level curves of $\epsilon$ in the plane ($a$, $m_1+m_2$), with the approximated upper limit ($\epsilon \sim $ 0.05)  shown with a dotted red line. The binary systems below the dashed blue curve cannot be approximated by the model, but these are very separated pairs of small masses. Finally.  we can quickly  estimate the limit of the integrable approximation for any binary system from the graph.

\subsection{Numerical approximation of the one-degree of freedom model}

The weak effect of the Galactic perturbation for the case $\epsilon < $ 0.05 can also be deduced from the motion equation of the complete three-degree of freedom model.

 The temporal evolution of the mean anomaly (eq. \ref{eq10}) is defined as
\begin{equation}
\dot{M} = \frac{\partial \mathcal{K}_0}{\partial L}+\frac{\partial \mathcal{K}_1}{\partial L}+\frac{\partial \mathcal{K}_2}{\partial L} \,  \rm{.} \label{eq11}
\end{equation}

\noindent The second term in the right side of the eq. (\ref{eq11}) is zero since it is independent of $L$, the first term is the mean motion of the astrocentric orbit of $m_2$ around $m_1$, and the last term is the perturbation of the Galaxy,

\begin{equation}
\dot{M} = n + \frac{\partial \mathcal{K}_2}{\partial L}  \,  \rm{,} \label{eq12}
\end{equation}

\noindent where

\begin{equation}
 \frac{\partial \mathcal{K}_2}{\partial L} = - 4 \Omega_G A_G x \frac{\partial x}{\partial L}  + \nu_G^2 z \frac{\partial z}{\partial L}   \,  \rm{,} \label{eq13}
\end{equation}

\noindent which can be written using a simplified form (eq. (\ref{eq9b})and (\ref{eq9c})), i.e.

\begin{equation}
 \frac{\partial \mathcal{K}_2}{\partial L} \sim  - \dfrac{\epsilon}{4} \sqrt{\dfrac{\mu}{a^3}}  (2 x_1 x_L - 5.77 z_1 z_L)  \,  \rm{.} \label{eq13b}
\end{equation}

\noindent Because we are considering $\epsilon < $ 0.05 the second term in the right part of eq. (\ref{eq12}) is smaller than the mean motion of the binary system. Therefore, the perturbation of the Galaxy is small enough to use the approximation $M \sim n t + \tau_n$ ($\tau_n$ is an arbitrary constant). So, $M$ circulates with a frequency $n$, and the semimajor axis has an almost constant evolution with a small amplitude.

The precession frequency of the ascending node is defined as

\begin{equation}
\dot{\Omega} = \frac{\partial \mathcal{K}_0}{\partial H}+\frac{\partial \mathcal{K}_1}{\partial H}+\frac{\partial \mathcal{K}_2}{\partial H} \,  \rm{.} \label{eq14}
\end{equation}

\noindent In this equation the first term of the right part is zero since ${\mathcal{K}}_0$ is independent of $H$, the second term is the motion of the binary system around the Galaxy, and the last term corresponds to the tidal perturbation of the Galaxy,

\begin{equation}
\dot{\Omega} = - 2 \Omega_G + \frac{\partial \mathcal{K}_2}{\partial H} \,  \rm{.} \label{eq15}
\end{equation}

\noindent where

\begin{equation}
 \frac{\partial \mathcal{K}_2}{\partial H} = - 4 \Omega_G A_G x \frac{\partial x}{\partial H}  + \nu_G^2 z \frac{\partial z}{\partial H}   \,  \rm{,} 
\end{equation}

\noindent which can be written using a simplified form (eq. (\ref{eq9b})and (\ref{eq9c})), i.e. 

\begin{equation}
 \frac{\partial \mathcal{K}_2}{\partial H} \sim  -  \dfrac{\epsilon}{4} \sqrt{\dfrac{\mu}{a^3}}   (2 x_1 x_H - 5.77 z_1 z_H)  \,  \rm{.} 
\end{equation}

\noindent In this case, for $\epsilon < $ 0.05 the second term in the right side of eq. (\ref{eq15}) is smaller than the angular velocity  of the binary system around the Galaxy (i.e.  $\Omega_G$).

Then, since the perturbation of the Galaxy is small we can write $\Omega \sim  2 \Omega_G t + \tau_o$ ($\tau_o$ is an arbitrary constant), which corresponds to a circulation of the angle $\Omega$ with frequency $2 \Omega_G$, the action $H$ has small variations and can be considered almost constant. The circulation of $\Omega$ is related to the orbital period of the binary system around the centre of the Galaxy.

The frequency of the argument of pericentre is defined as

\begin{equation}
\dot{\omega} = \frac{\partial \mathcal{K}_0}{\partial G}+\frac{\partial \mathcal{K}_1}{\partial G}+\frac{\partial \mathcal{K}_2}{\partial G} \,  \rm{.} \label{eq17}
\end{equation}

\noindent The first and second terms on the right side are independent of $G^*$ and then are equal to zero. The last term, which is small, defines the tidal perturbation of the Galaxy on the binary system.  So, the frequency is defined by
\begin{equation}
\dot{\omega} = \frac{\partial \mathcal{K}_2}{\partial G} = - 4 \Omega_G A_G x \frac{\partial x}{\partial G}  + \nu_G^2 z \frac{\partial z}{\partial G}  \,  \rm{,} \label{eq18}
\end{equation}

\noindent  which can be simplified with the approximations (\ref{eq9b}) and (\ref{eq9c}),  

\begin{equation}
 \frac{\partial \mathcal{K}_2}{\partial G} \sim  -  \dfrac{\epsilon}{4} \sqrt{\dfrac{\mu}{a^3}}   (2 x_1 x_G - 5.77 z_1 z_G)  \,  \rm{,} \label{eq17b}
\end{equation}

\noindent which define the secular evolution of the binary system.

Thus, for $\epsilon < $ 0.05 the equations of motion for the original three-degree of freedom model can be approximated by 
\begin{equation}
\begin{array}{cclccl}
\dot{M} &\sim& n , & \dot{L} &\sim& 0 \,  \rm{,} \\ 
   \\
\dot{\omega} &=& \dfrac{\partial \mathcal{K}_2}{\partial G} , & \dot{G} &=& -\dfrac{\partial \mathcal{K}_2}{\partial \omega} \,  \rm{,} \\ 
   \\
\dot{\Omega} &\sim& -2 \Omega_G , & \dot{H} &\sim& 0 \,  \rm{.} \label{eq19}
\end{array}
\end{equation}

\noindent In this approximation $L$ and $H$ are almost constants with small variations, and the angles $M$ and $\Omega$ circulate with an osculating period $T_n \sim 2 \pi / n$ and a Galactic rotational period  $T_{o} \sim  \pi / \Omega_g \sim 10^8 yrs $,  respectively. The latter defines the small precession of the ascending node, which is related to the circulation of the system around the Galaxy. Hence, the tidal secular evolution has a dynamical behaviour that can be approximated by a system of one-degree of freedom and is defined by the pair action-angle variables  $G$ and $\omega$. The almost constant value of $H$ implies  a dynamical behaviour of $e$ and $I$ that is similar to the dynamic of the Lidov-Kozai resonance, such as we deduce from the averaging process. 

\subsubsection{Frequency analysis}

From the three natural frequencies involved in the problem   we can perform a brief comparative analysis between them, such as that carried out by Antognini (2015). A frequency analysis  (Michtchenko et al. 2002) is very important because this allows us to predict the interaction between the natural frequencies of the system, which can act as a source of chaos (i.e. commensurable periods). Our previous deduction of the angular frequencies allows us to define the secular ($f_s$), node-precession ($f_n$), and osculating ($f_o$) frequencies as
\begin{equation}
\begin{array}{ccl}
f_s = \dot{\omega} &\propto& \Omega_G A_G \sqrt{\dfrac{a^3}{\mu}} \,  \rm{,} \\ 
   \\
f_n = \dot{\Omega}  &\propto& 2 \Omega_G \,  \rm{,} \\
\\
f_o = \dot{M} &\propto& \sqrt{\dfrac{\mu}{a^3}} \,  \rm{.} \\ 
  \label{eq19r}
\end{array}
\end{equation}

\noindent Then, the ratio between these frequencies are

\begin{equation}
\begin{array}{ccl}
f_o/f_s  &\sim&  \dfrac{4}{\epsilon } \,  \rm{,} \\ 
   \\
f_n/f_s  &\sim& 4 \sqrt{\dfrac{\Omega_G}{ A_G \epsilon }}   \,  \rm{,} \\
\\
f_o/f_n  &\sim&  \sqrt{\dfrac{A_G}{ \Omega_G \epsilon }} \,  \rm{.} \\ 
  \label{eq19rx}
\end{array}
\end{equation}

  \begin{figure}
\centering
\includegraphics [width=0.45\textwidth] {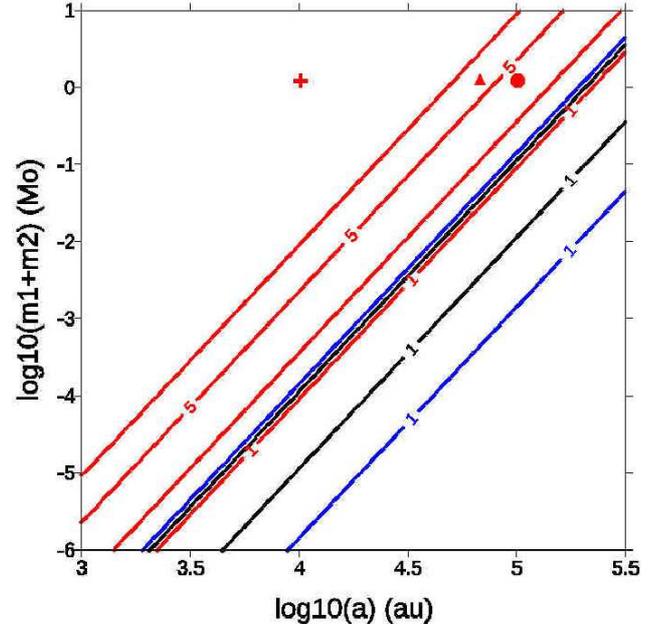} 
\caption{{\small Level curves of the ratio between the frequencies: $f_o/f_s$ (dashed black line),  $f_n/f_s$ (dotted blue line) and $f_o/f_n$ (red line). For the comparison with the secular frequency we plot the ratio 1 and 10. For the comparison between the fast frequencies ($f_o/f_n$) we plot the levels 1, 2, 5 and 10. The red symbols (cross, triangle and circle) represent the position of three binary systems (see Sect. \ref{standard}). }}
\label{figxx}
\end{figure}

Figure \ref{figxx} shows the ratio between the frequencies for the same range of masses and semimajor axis of Fig. \ref{figx}. The relation between the high frequencies and the secular frequency are  indicated with a dashed black line for $f_o/f_s$, and dotted blue line for $f_n/f_s$. For simplicity, we only plot levels 1 and 10 because the commensurable region is restricted to  binary systems with particular characteristics. The ratio $f_o/f_n$ is plotted with a red line and we show the levels 1, 2, 5, and 10. For the high frequencies, the commensurability ($f_o/f_n \leq 5$) achieves a greater region in the plane ($a$, $m_1+m_2$).

For binary systems with $a < 3 \times 10^6$ au ($\sim$ 1.5 pc), the problem of the interaction of the secular frequency with the other periods is restricted to small masses. For binary systems with $m_1+m_2 >$ 0.1 M$_\odot$, $f_o$ and $f_n$  are commensurate for  $a >$ 60000 au. Then, the application of the secular one-degree of freedom model is accurate  for binary systems with $f_o/f_n \geq$ 5. The scope of the model (Sect. \ref{scope}) is close to the start of the non-secular region.

Finally, the approximated secular frequency for binary systems with $a<$  1000 au  and $m_1+m_2 <$ 10 M$_\odot$ (eq. \ref{eq19r}) has an associated period of hundred of Gyr. Such a period is an order of magnitude greater than the age of the universe (i.e. $\sim$ 13 Gyr). This gives a theoretical explanation for the empirical  limit of 1000 au between tight and wide binary systems .

  \begin{figure}
\centering
\includegraphics [width=0.4\textwidth] {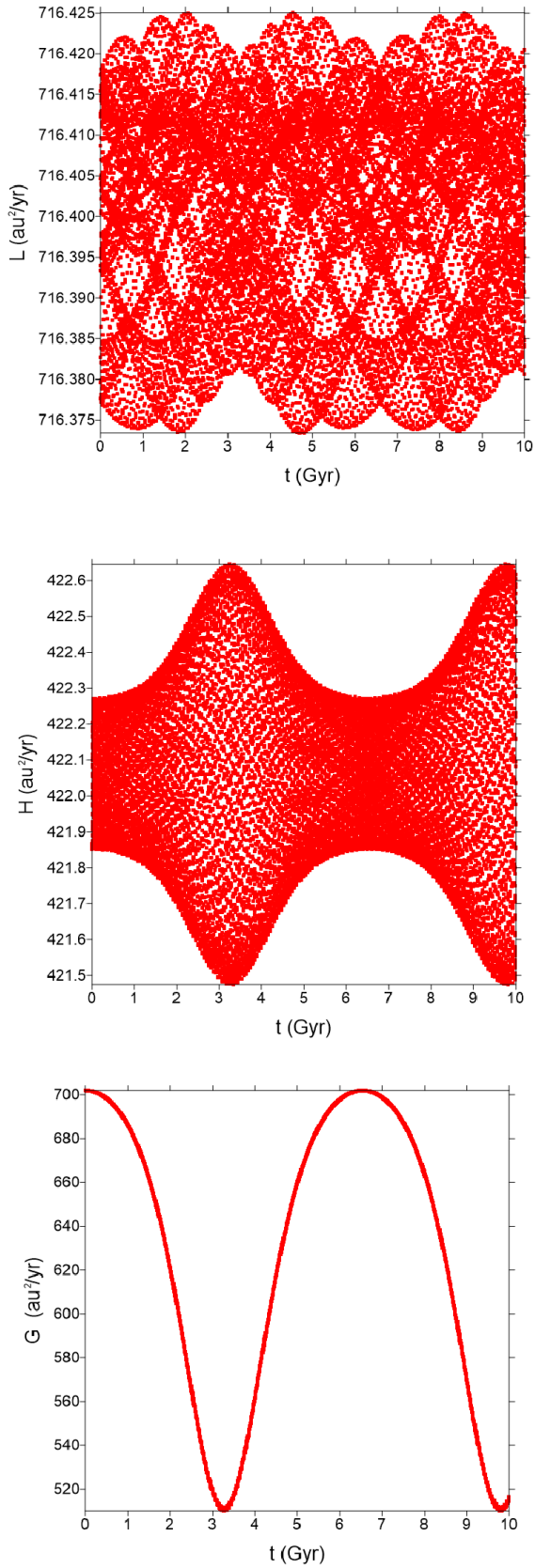} 
\caption{{\small Time evolution of the action variables of Delaunay $L$ (top panel), $H$ (middle panel) and $G$ (bottom panel) for our \textit{standard system}. The small variations of $L$ and $H$ and the circulation of $M$ and $\Omega$ (Fig. \ref{fig1b}) confirm the proposed approximation.}}
\label{fig1a}
\end{figure}

 \begin{figure}
\centering
\includegraphics [width=0.4\textwidth] {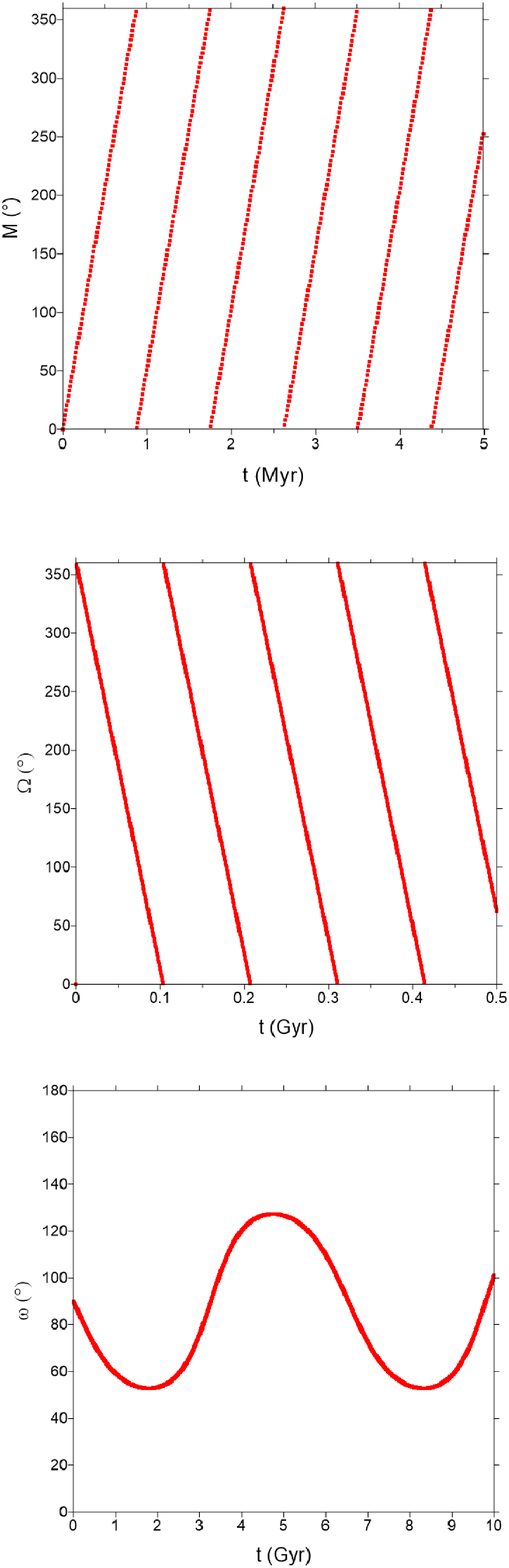}
\caption{{\small Time evolution of the angular variables of Delaunay $M$ (top panel), $\Omega$ (middle panel) and $\omega$ (bottom panel) for our \textit{standard system}. The small variations of $L$ and $H$ (Fig. \ref{fig1a}) and the circulation of $M$ and $\Omega$  confirm the approximation proposed.}}
\label{fig1b}
\end{figure}

\subsection{Testing the one-degree of freedom model: An application example}\label{standard}

 \begin{figure}
\centering
\includegraphics [width=0.45\textwidth] {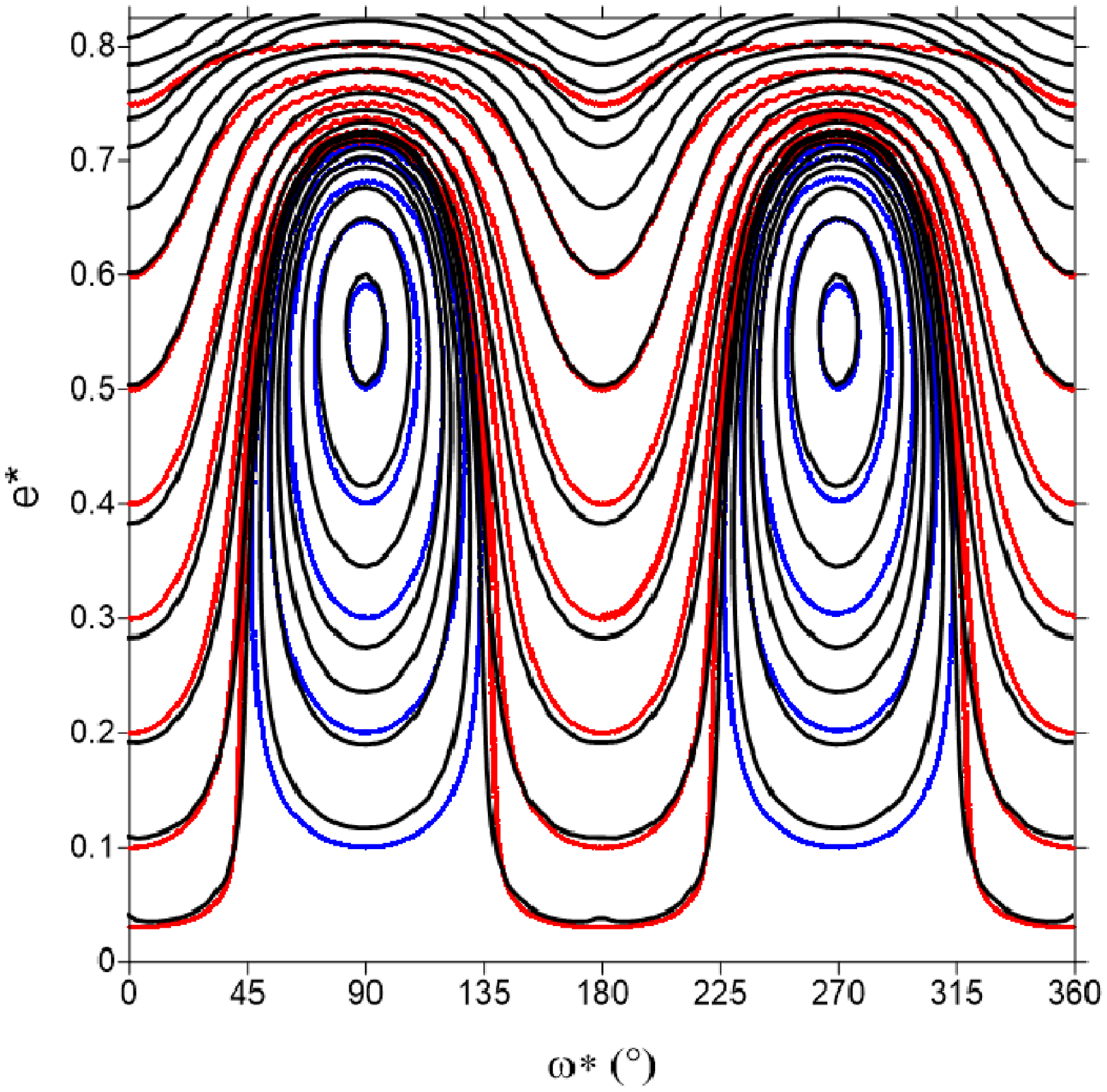}
\caption{{\small Comparison between the predictions of our model (dashed black line) and numerical simulations (red and blue lines) in the plane ($\omega^*, e^*$), for the family of solutions of the \textit{standard system}.}}
\label{fig2}
\end{figure}

In order to show the use of the model we use a binary system with masses $m_1=$ 1.0 M$_\odot$ and $m_2=$ 0.3 M$_\odot$ as an example. The secondary star is moving around the main star in an initial configuration named "\textit{standard system}", where the initial parameters are $a_0=$ 10000 au ($L_0=$ 716.4 au$^2$/yr), $e_0=$ 0.2 ($G_0=$ 702 au$^2$/yr), $I_0=$ 53$^\circ$ ($H_0=$ 421.86 au$^2$/yr), and $\omega_0=$ 90$^\circ$. The other two angles are equal to zero. Figures \ref{fig1a} and \ref{fig1b} show the temporal evolution of the action and angle variables of Delaunay, respectively. In these figures the simulations were carried out by integrating the exact equations of motion through a Bulirsch-Stoer code with adopted accuracy of $10^{-13}$.

On the top graphs of each Figure we show the evolution of the osculating elements, the middle panels show the evolution of motion in the Galactic $z$ direction, and the bottom panel of each figure represents the secular motion. In the case of $M$, there is a circulation and $L$ has a small variation ($\Delta L/L_0 \sim 10^{-4}$). The same behaviour is followed by the angle $\Omega$ which circulates, and the variation of $H$ is also small ($\Delta H/H_0 \sim 10^{-3}$). If we consider the secular motion, it is evident that the influence of the two first degrees of freedom considered are small and this confirms that $\dot L^* \sim 0$ and $\dot H^* \sim  0$, and then the simple model is useful in this case.

On the other hand it is remarkable that the tidal secular period is long ($\sim 7 \times 10^9$ yrs) and has a timescale that is comparable to the age of the solar system as we can see in the evolution of $G$ and $\omega$. The angle $\omega$ librates (or oscillates) and later we determine if this motion is resonant or not. The period of the precession of the ascending node is $\sim 10^8$ years, and it is seen in the pair angle-action $\Omega$ and $H$. This dynamical behaviour shows a combined dynamics between the tidal secular component and the precession of the stellar pair, but owing to the great differences between the terms in the Hamiltonian  ($ \mathcal{K}_1 \gg \mathcal{K}_2$), we can approximate the decoupling of both dynamical components. This is also valid for the osculating behaviour, where the secular and node precession are responsible for the amplitude different to zero in $L$. The limits for such approximation are determined in the following sections.

However, our \textit{standard system} (Fig. \ref{fig1a} and \ref{fig1b}) is only one case. In order to show a more robust confirmation of our approach, we considered a set of binary systems with equal values of masses, $H^*$ and $L^*$. We call such set of solutions a \textit{family}. The mean-mean variables, which are needed in the one-degree of freedom model, can be obtained with the generatrix function. Nevertheless, because of the small amplitude of the  two osculating actions variables $L$ and $H$, we can use the initial values to approximate their mean-mean values  $L^*$ and $H^*$. 

Although, the action $G^*$ is a canonical variable, for practical applications the eccentricity is more useful.  Fig. \ref{fig2} shows, with a dashed black line, the level curves of constant $\mathcal{K}^*$ for the family of the \textit{standard system} (i.e. equal values of $m_1$, $m_2$, $L^*$, and $H^*$) in the plane ($\omega^*, e^*$), which were calculated with the one-degree of freedom model. In Figure \ref{fig2} we also include the numerical simulation of ten binary systems, with the same initial values of $H$ and $L$, that could be approximated by the mean-mean values. In that figure the binary systems with circulation in the angle $\omega$ are indicated in red, and those with oscillation are shown in blue. We can see that our model is a good approximation of the real system because the other two degrees of freedom have a small influence in the secular dynamic of the real binary system, and this allows us to approximate the mean-mean variables $L^*$ and $H^*$ by their osculating variables $L$ and $H$. 

However, there is a small difference between the model and numerical simulations at high eccentricities in Fig. \ref{fig2}  because of our assumption that the osculating actions are similar to their averaged value. Although in a first approximation the prediction can be useful, for a more accurate result it is necessary to calculate the averaged value of $L$ and $H$ with the generating function $\chi_g$ or with a short numerical integration.

 \begin{figure}
\centering
\includegraphics [width=0.45\textwidth] {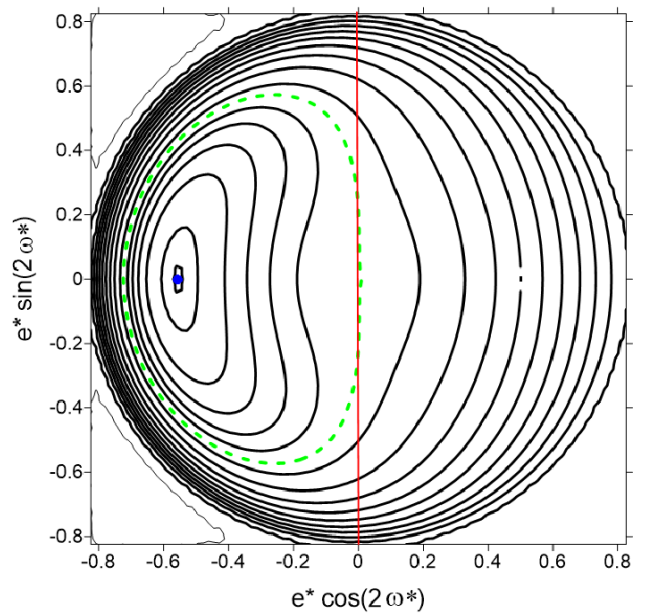}
\caption{{\small Energy levels of the secular Hamiltonian (\ref{eq25}) in the plane ($e^* \cos{2 \omega^*}$ - $e^* \sin{2 \omega^*}$) for the family of solutions of the \textit{standard system}. The dashed green line shows the kinematic transition from oscillation to circulation.}}
\label{fig3}
\end{figure}

\begin{figure}
\centering
\includegraphics [width=0.4\textwidth] {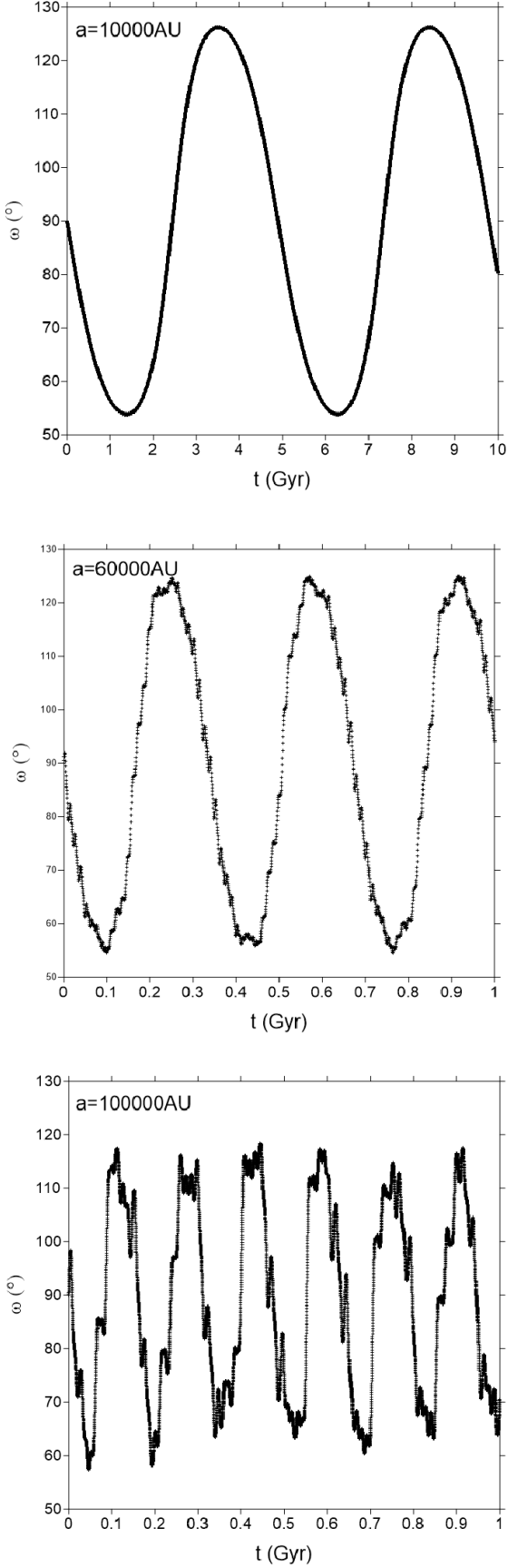}
\caption{{\small Evolution of the angle $ \omega$ for the binary system of Fig. \ref{fig1a} and \ref{fig1b}, but with the following three different values of semimajor axis: $a=10000$ au (top panel),  $a=60000$ au (middle panel), and $a=100000$ au (bottom panel).}}
\label{fig9a}
\end{figure}

\begin{figure}
\centering
\includegraphics [width=0.45\textwidth] {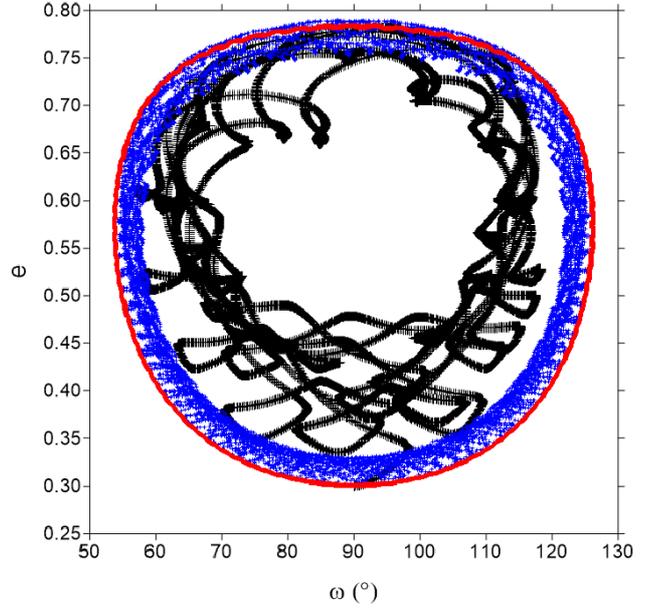}
\caption{{\small Dynamical evolution of the the binary system of Fig. \ref{fig1a} and \ref{fig1b} in the plane ($\omega,e$), but for the following three different values of semimajor axis: $a=10000$ au (red),  $a=60000$ au (green), and $a=100000$ au (black).}}
\label{fig9}
\end{figure}

Moreover, in order to test our analytic estimations of the quasi-integrable approximation and the ratio of frequencies, we carry out three numerical simulations with different semimajor axes. We consider the \textit{standard system} with the following different separations:  $a=10000$ au  ($\sim 0.05$ pc), $a=60000 $ au (i.e. $\sim 0.3 $ pc), and $a=100000$ au ($\sim 0.5$ pc). In
Figs. \ref{figx} and \ref{figxx}, the three cases are indicated with red symbols as follows: a cross for $a=10000$ au, a triangle for $a=60000$ au, and a circle for $a=100000$ au.  For the three examples, we plot the temporal evolution of the angle $\omega$ and the dynamical behaviour in the plane ($\omega, \, e$) in Figures \ref{fig9a} and  \ref{fig9},  respectively. For $a=60000$ au the temporal evolution shows other  perturbations in addition to the secular frequency and for larger semimajor axes (i.e. $a>60000$ au) our approximation is no longer valid (i.e. $f_o$ and $f_n$ are commensurable), which agrees with our analytical deduction.  

The prediction of the secular one-degree of freedom model shows a good agreement with the numerical experiments for $\epsilon<$ 0.05 and out of the region of commensurable frequencies, indicating that the model seems to be accurate enough for the approximations considered.  

\section{Stationary solutions of the averaged Hamiltonian}\label{num}

The family of solutions in Figure \ref{fig2} shows two stable stationary solutions at ($\omega^*, e^*$) $\sim$ (90$^\circ$, 0.54) and (270$^\circ$, 0.54). Such points are equilibrium solutions (maxima or minima) of the tidal secular dynamic defined by the Hamiltonian function. From  the mean-mean Hamiltonian of the one-degree of freedom model, we can find the position of the equilibrium solutions and also we are able to determine if the coupling between $e^*$ and $I^*$ ($H^* =$ cte) is a consequence of the Lidov-Kozai type resonance or not. In the case of a resonant behaviour there are three equilibrium points, while in a non-resonant one there is a single point (Murray \& Dermott 1999).

The averaged mean-mean Hamiltonian is
\begin{equation}
\mathcal{K^*} = - \dfrac{\mu^2}{2{L^*}^2} - 2 \Omega_G {H^*} + \epsilon \mathcal{K}^*_{g1} ({G^*},{\omega^*};{L^*},{H^*}) \,  \rm{,} \label{eq25}
\end{equation}

\noindent where the two first terms are constant (i.e. $H^*$ and $L^*$), and the equilibrium points of the Hamiltonian are defined by the last term. Thus, our function is

\begin{equation}
\mathcal{K}^*_{g1} = \mathcal{K}^*_{g1} ({G^*},{\omega^*};{L^*},{H^*}) \,  \rm{,} \label{eq25b}
\end{equation}

\noindent with variables ${G^*}$ and ${\omega^*}$. 

The zero of the equations of motion are

\begin{equation}
\begin{array}{cclccl}
   \\
\dot{\omega^*} &=& \epsilon \dfrac{\partial \mathcal{K}^*_{g1}}{\partial G^*}  & = & 0 \,  \rm{,} \\
\\
 \dot{G^*} &=& -\epsilon \dfrac{\partial \mathcal{K}^*_{g1}}{\partial \omega^*} & = & 0 \,  \rm{,} \\ 
   \\
 \label{eq26}
\end{array}
\end{equation}

\noindent where the frequency of $G^*$ is the derivative of the Hamiltonian with respect to the angle  $\omega^*$, and only the second term is a function of this angle (Equation \ref{eq61}). Then, the equilibrium solutions are possible for $\sin{2\omega^*}=0$ (i.e. ${2\omega^*}=$ $m \pi$, with $m=$ 0, 1, 2, etc.).

On the other hand, the frequency of $\omega^*$ is the derivative of the Hamiltonian with respect to the action $G^*$,

\begin{equation}
\begin{array}{ccl}
 \dfrac{\partial \mathcal{K}^*_{g1}}{\partial G^*}  &=&  \dfrac{\mu^{2}}{4\,{L^*}^2} \bigg\lbrace  G^* [\alpha (2 \cos{2 \omega^*} - 1 ) + 1 ] \\
 &&\\
 &-&  \dfrac{ 2 \alpha  (\cos{2 \omega^*} - 1) {H^*}^2 {L^*}^2 }{{G^*}^{3}} \bigg\rbrace  \,  =  \,  0  \,  \rm{,}\\
 \label{eq27}
\end{array}
\end{equation}

\noindent and solving for  $G^*$ we obtain

\begin{equation}
\begin{array}{ccl}
{G^*}^4 & = & \dfrac{{H^*}^2 {L^*}^2 (\cos{2 \omega^*} - 1)}{ \cos{2 \omega^*} - 0.5 + ( 2 \alpha)^{-1}} \,  \rm{,} \\
 \label{eq28}
\end{array}
\end{equation}

\noindent with $( 2 \alpha)^{-1} \sim 0.1477 $.

For an even multiple of $\pi$ (i.e. $ m \pi $, where  $m=$ 0, 2, 4, etc.), the term $\cos{2 \omega^*} - 1$ is always $0$. Hence,  $G^*=0$ ($e^*=1$) and the solutions are in the limit of an hyperbolic motion. Moreover, because the coupling of $e$ and $I$ (i.e. $H^* =$ constant), there is a single value of $H^*$ ($H^* =$ 0) that satisfies such solutions. Therefore, we can discard these equilibrium points because they are not a general solution of each family and their study are out the scope of our study.

In the case of odd values of $m$  the equilibrium points are

\begin{equation}
\begin{array}{cclccl}
{G^*_M}^2 & = & {H^*} {L^*} \sqrt{\dfrac{-2}{ - 1.5 + ( 2 \alpha)^{-1}}} \, \rm{,} & 2\omega^* &=& m \pi \, (m=1  , 3)  \rm{,}\\
\\
{G^*_M} & \sim & 1.103 \sqrt{{H^*} {L^*} } \, \rm{,} & \omega^* &=& m \pi/2 \, (m=1, 3)   \rm{.}
 \label{eq29}
\end{array}
\end{equation}

\noindent Considering the \textit{standard system} of Section \ref{standard}, we found $G^*_M \sim 606$ au$^2$/yr, or $e^*_M \sim 0.54$ and $\omega^*=$ 90 and 270$^\circ$ ($2\omega^*=$ 180$^\circ$), which is coincident with the maxima shown in the Figure \ref{fig2}.

Therefore, we found that for each combination of $H^*$ and $L^*$ there is a single equilibrium point at $2\omega^*=$ 180$^\circ$, whose value $e^*_M$ in eccentricity could be interpreted as a "forced eccentricity" ($e_f$). The forced eccentricity is a fixed value and the curves around  it (e.g. $e_f$) could be defined as a proper or free eccentricity with their own free frequency.

The existence of a single maximum indicates a similar secular (non-resonant) motion, even if the coupling between the eccentricity and the inclination seems to behave like the Lidov-Kozai resonance, but the dynamical behaviour is different because there is not a separatrix. The secular motion without resonance could be better appreciated in the plane of the regular variables  $x=e^* \cos{(2 \omega^*)}$ and  $y=e^* \sin{(2 \omega^*)}$, where the motion is defined as quasi-concentric circles around the forced eccentricity ($e^*_M$). The level curves of constant mean-mean Hamiltonian of Figure \ref{fig2} (i.e. family of the \textit{standard system}) are repeated in the new plane ($x,y$) in Figure \ref{fig3}, where the absence of a separatrix is evident. In fact, we have  quasi-concentric circles with the centre separated from the origin.

 \begin{figure}
\centering
\includegraphics [width=0.45\textwidth] {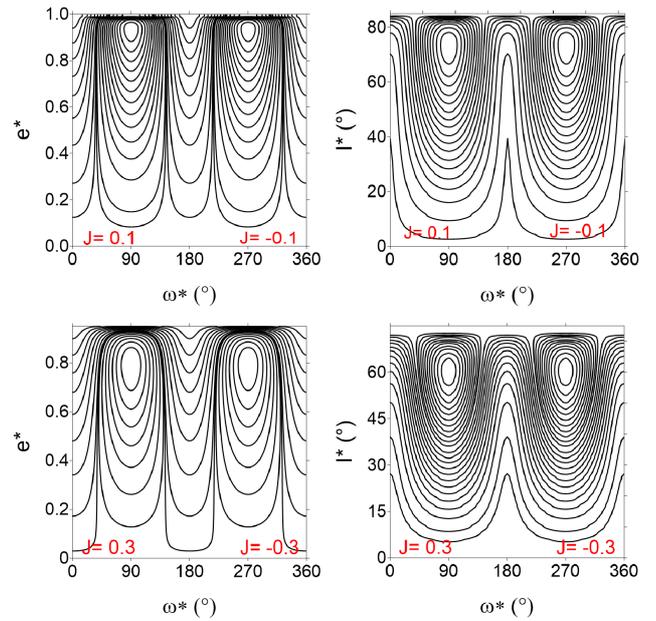}
\caption{{\small Structure of the secular phase space of the \textit{standard system} for the following four different values of the DPAM ($J=H^*/L^*$): 0.1 and -0.1 (top panels), and 0.3 and -0.3 (bottom panels). The left panels show the ($\omega^*-e^*$) plane, and the right panels show the ($\omega^*-I^*$)-plane. Although the dynamical structure is similar, the position of the equilibrium points in the $e^*$-axis (and $I^*$-axis) and the extension of the phase space decrease with the increase of $J$. }}
\label{fig4a}
\end{figure}

The difference between the dynamical structures of  resonant and  non-resonant motion allows us to understand why this Hamiltonian is only secular (Murray \& Dermott 1999). Indeed, although the angles could oscillate in both cases, the oscillations are topologically different. In the resonant case the transition of the angle from oscillation to circulation occurs through true bifurcations of the solutions, along the separatrix that contains a saddle-like point. In other words, the two regimes of motions, oscillations, and circulation are topologically distinct. In this case, we can say that the resonant angle librates and the system is in a true resonance state. In the other case (our study case) the oscillations are merely kinematic since all levels, even those passing through the origin (green dashed curves), belong to the same structurally stable family. In other words, there is no topological difference between oscillations and circulations. 

Finally, a two-body system that is disturbed by a third body is able to develop a resonance of Lidov-Kozai (Lidov 1961; Kozai 1962; Antognini \& Thompson 2016). In this work we are dealing with a similar problem, where a binary system is disturbed by an external force, but we found a secular coupling of $e$ and $I$. The difference seems to be in the form of the third body; in three star systems the disturbing body is a point mass, while we are working with a potential corresponding to an extended body (i.e. the Galaxy).

 \begin{figure}
\centering
\includegraphics [width=0.45\textwidth] {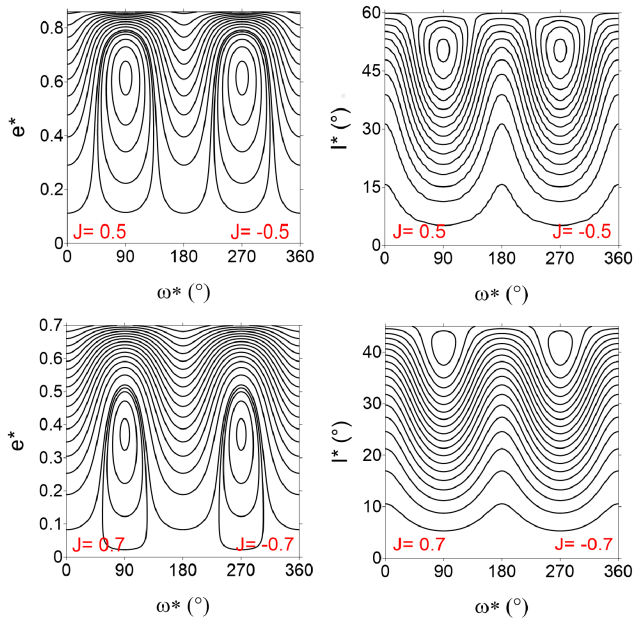}
\caption{{\small Same as in Fig. \ref{fig4a} except for DPAM ($J=H^*/L^*$): 0.5 and -0.5 (top-panels), and 0.7 and -0.7 (bottom-panels). }}
\label{fig4b}
\end{figure}

\section{Dynamical portait of the secular evolution}\label{fase}

In the previous section we analysed the energy levels of a particular set of values for $H^*$ and $L^*$, which we named a "family", and now we consider other families to provide a complete description of the dynamical portrait of the problem. Since $H^* = L^* \sqrt{1-{e^*}^2} \cos{I^*}$, a better option is to work on fixing the value of $L^*$, which allows us to introduce the variable $J=H^*/L^*=  \sqrt{1-{e^*}^2} \cos{I^*}$ (i.e. a dimensionless projection of the angular momentum, DPAM), which varies between -1 to 1. 

Then, for the masses and initial value of $L$ of the \textit{standard system}, we chose eight different values for DPAM: $J=$ -0.7, -0.5, -0.3, -0.1, 0.1, 0.3, 0.5, and 0.7, which are represented in Figs. \ref{fig4a} and \ref{fig4b}. However, the Galactic potential is symmetric in the $z$-axis; in fact  the equilibrium points (eq. \ref{eq28}) and the equations of motion (\ref{eq26}) have a quadratic dependence on  $H^*$.  Thus, the secular phase space of  the binary system depends on the absolute value of $H^*$ or $J$. Therefore, we represent the eight DPAM in four  dynamical maps , one for each positive-negative pair of $J$. In order to make a more complete dynamical analysis, we represent each map in two planes: the ($\omega^*-e^*$)-plane and ($\omega^*-I^*$)-plane.

We can see that the general structure of the phase space does not change and there are two maxima for mean-mean Hamiltonian at $\omega^*= 90^\circ$ and 270$^\circ$. The most important differences between the eight maps are the following: first, the position of the equilibrium points of the Hamiltonian, which decreases in eccentricity (and $I^*$) when $J$ increase; second, smaller values of $J$ allow greater values of eccentricity (and $I^*$) and the extension of the phase space increase in the $e^*$ direction ($I^*$ direction). We are working with mean-mean variables, but we choose the initial value of $L$ as our mean-mean value $L^*$, which is possible because of the small amplitude of the action variables $L$ and $H$ (see Sect. \ref{standard}).

\begin{figure}
\centering
\includegraphics [width=0.4\textwidth] {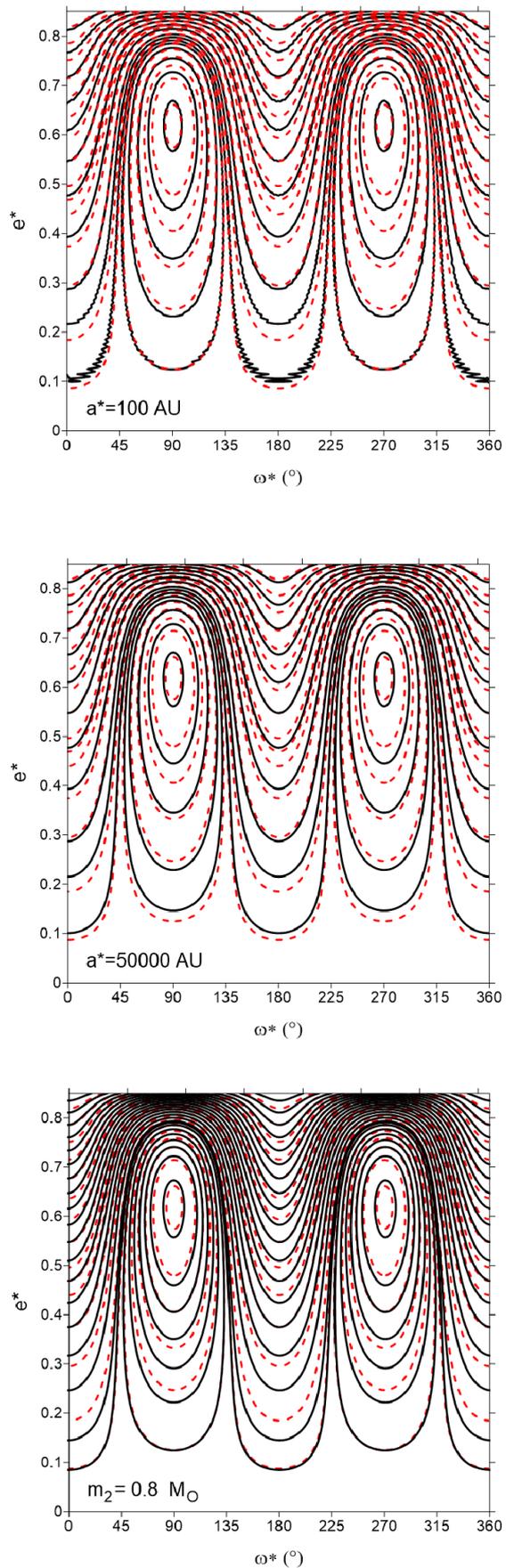}
\caption{{\small Phase space for different values of  $L^*$: $a^*$ = 100 (top panel), 50000 au (middle panel), and  $m_2=$ 0.8 M$_\odot$ (bottom panel). As comparison the levels of constant Hamiltonian for $a$ = 10000 au of the Fig. \ref{fig4b} (top panel) are indicated with dashed red line. }}
\label{fig5}
\end{figure}

Next, we changed the value of $L^*$, keeping fixed $J$ ($H^*/L^*$) in $0.5$ (Fig. \ref{fig4b}, top panel). However, $L^*$ depend on the masses and the semimajor axis, and then we choose three test cases with $a$ = 100 and 50000 au, and  $m_2=$ 0.8 M$_\odot$. Figure \ref{fig5} shows our results (black line), where the levels of constant mean-mean Hamiltonian for $a$ = 10000 au (i.e. Fig. \ref{fig4b}, top panel) were included for comparison (dashed red line). The dynamical structure is equal in the three cases, and it agrees very well with the example of reference (dashed red lines).

The independence of the dynamic structure with the variable $L^*$ is a very important result, which indicates that the long-period dynamics of a binary system in the Galaxy depends on the $z$ component of the angular momentum ($H$), but it is independent of the semimajor axis and the masses ($L$). Because of such independence of the masses, the dynamical behaviour for any binary system (i.e. star-star, Sun-comet, etc.) is always the same for similar values of DPAM ($J$).

Such results can be analytically deduced from the mean-mean Hamiltonian,

\begin{equation}
\mathcal{K^*} = - \dfrac{\mu^2}{2{L^*}^2} - 2 \Omega_G {H^*} +  \epsilon \mathcal{K}^*_{g1} ({G^*},{\omega^*};{L^*},{H^*}) \, \rm{.} \label{eq30}
\end{equation}

\noindent The two first terms of the Hamiltonian are constants, thus solving for $\mathcal{K}^*_{g1}$ we can define a new Hamiltonian,

\begin{equation}
\mathcal{K}_J  \, =  \, \epsilon \dfrac{\mu^{2}}{4\,{L^*}^2}  [ \alpha (\sin{{I^*}})^2  (1+{e^*}^2 -   {e^*}^2 \cos{2\omega^*}) -  (1+{e^*}^2)  ]  \,  \rm{.} \label{eq31}
\end{equation}

\noindent Since $\epsilon {L^*}^{-2}{\mu}^{2}/4$ is a constant,  we can move it to the left side and we obtain a new Hamiltonian, which is only function of $e^*$ and $I^*$ , and is written as

\begin{equation}
\mathcal{K}_{J2} = \alpha (\sin{{I^*}})^2  (1+{e^*}^2 -   {e^*}^2 \cos{2\omega^*}) -  (1+{e^*}^2)  \, \rm{.} \label{eq32}
\end{equation}

\noindent Then, although the absolute value of the Hamiltonian $\mathcal{K}_{J2}$ changes for different values of $a^*$ or the masses, its functional form depends only on the eccentricity and inclination, which explains the agreement in Figure \ref{fig5}.

However, the influence of the Galactic potential is only reported for wide binary systems (Brunini 1995; Jiang \& Tremaine 2010), and it is ignored for tight binary systems ($a<$ 1000 au). Therefore we could explore these systems looking for the dependence of the potential with the separation. We find that the difference lies in the period of the tidal secular cycle. From the equation (\ref{eq30}), the Hamilton equations are


\begin{equation}
\begin{array}{ccl}
   \\
\dot{\omega^*} = \epsilon \dfrac{\partial \mathcal{K}^*_{g1}}{\partial G}  &=& \epsilon \dfrac{\mu^{2}}{4\,{L^*}^2}  \bigg\lbrace  G^* [\alpha (2 \cos{2 \omega^*} - 1 ) + 1 ] \\
 &&\\
 &-&  \dfrac{ 2 \alpha  (\cos{2 \omega^*} - 1) {H^*}^2 {L^*}^2 }{{G^*}^{3}} \bigg\rbrace  \, \rm{,} \\

   \\

\dot{G^*} = - \epsilon \dfrac{\partial \mathcal{K}^*_{g1}}{\partial \omega}  &=& - \epsilon \dfrac{2 \alpha \mu^{2}}{4\,{L^*}^2\,{G^*}^2}  ({L^*}^2 - {G^*}^2) ( {G^*}^2 -{H^*}^2) \sin{2 \omega^*} \, \rm{.}  \\
 \label{eq33}
\end{array}
\end{equation}

\noindent Once again, working algebraically we obtain

\begin{equation}
\begin{array}{ccl}
   \\
\dot{\omega^*}  &=& \Omega_G A_G \dfrac{{L^*}^3}{\mu^{2}} \bigg\lbrace  \sqrt{1-{e^*}^2} [\alpha (2 \cos{2 \omega^*} - 1 ) + 1 ] \\
 &&\\
 &-&  \dfrac{ 2 \alpha  (\cos{2 \omega^*} - 1) {(\cos I^*)}^2  }{\sqrt{1-{e^*}^{2}}} \bigg\rbrace  \,  \rm{,}\\
 && \\
\dot{G^*}   &=& - 2 \Omega_G A_G  \alpha \dfrac{{L^*}^4}{ \mu^2} {e^*}^2 (\sin I^*)^2 \sin{(2 \omega^*)} \, \rm{.}  \\
 \label{eq34}
\end{array}
\end{equation}

Thus, the functional form of the evolution of the variables $\omega^*$ and $G^*$ is the same and similar to the mean-mean Hamiltonian, it is independent of the semimajor axis and the masses. However, the terms ${L^*}^3 \mu^{-2}$ and ${L^*}^4 \mu^{-2}$ in $\dot{\omega^*}$ and $\dot{G^*}$, respectively, change the frequency of the secular cycle. It is worth to note that, the evolution of $\omega^*$ is dimensionless, but $\dot{G^*}$ is parametrized by  $L^*$. So, the motion equations $\dot{\omega^*}$ and $\dot{G^*}{L^*}^{-1}$ are a better choice to understand the dynamical difference between binary systems.

For the modified motion equations ($\dot{\omega^*}$ and $\dot{G^*}/L^*$), the tidal secular period is defined by ${L^*}^3 \mu^{-2}$ (${a^*}^{1.5} \mu^{-0.5}$).  These results allow us to understand the difference between tight and wide binary systems with similar masses. The dynamical behaviour is the same for any binary system, but the time required to complete the secular cycle changes as a function of the separation and the masses of the components of the pair.

 \begin{figure}
\centering
\includegraphics [width=0.45\textwidth] {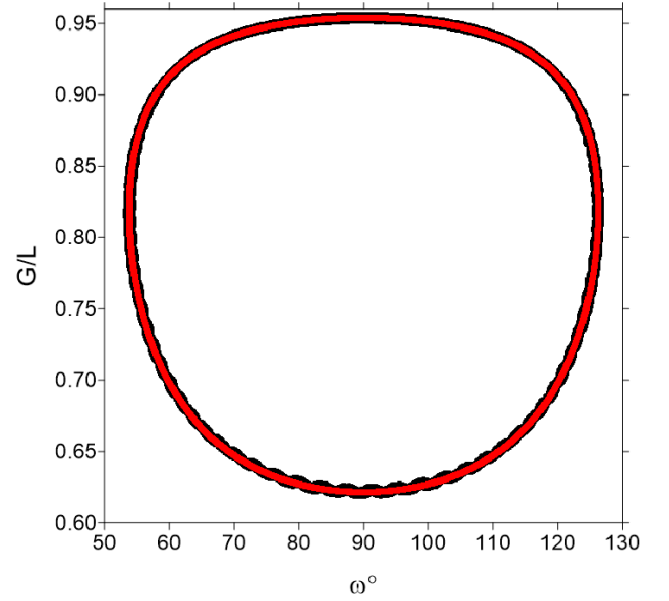}
\caption{{\small Secular phase space of the \textit{standard system} for $a=$ 10000 au (black line) and $a=$ 1000 au (red line), in the plane $(\omega,G/L)$. Because we find good agreement between the two binary systems, we decided to reduce the size of the red curve to appreciate the black curve.}}
\label{fig6a}
\end{figure}

The analytic result deduced from eq. (\ref{eq34}) is illustrated by the \textit{standard system} of Section \ref{standard}. For the binary star with masses $m_1 =$ 1.0 M$_\odot$ and $m_2 =$ 0.3 M$_\odot$, we consider two configurations:  $a_A=$ 10000 au and $a_B=$ 1000 au. From our one-degree of freedom model, in the modified motion equations ($\dot{\omega^*}$ and $\dot{G^*}/L^*$) we can predict a relation between the secular periods as

\begin{equation}
\begin{array}{ccccccl}
   \\
\dfrac{\dot{\omega^*_B}}{\dot{\omega^*_A}} &=& \dfrac{T_A}{T_B}  &\sim& ({a^*_B}/{a^*_A})^{1.5} &\sim& 0.031 \,  \rm{,}\\
\\
\dfrac{\dot{G^*_B}L^*_A}{\dot{G^*_A}L^*_B} &=& \dfrac{T'_A}{T'_B}  &\sim& ({a^*_B}/{a^*_A})^{1.5} &\sim& 0.031 \,  \rm{.} \\
 \label{eq35}
\end{array}
\end{equation}

\noindent Then, the wide binary star (10000 au) has a secular period  $\, \sim$ 31 times smaller than the tidal period of the more compact binary star (1000 au). 

\begin{figure}
\centering
\includegraphics [width=0.45\textwidth] {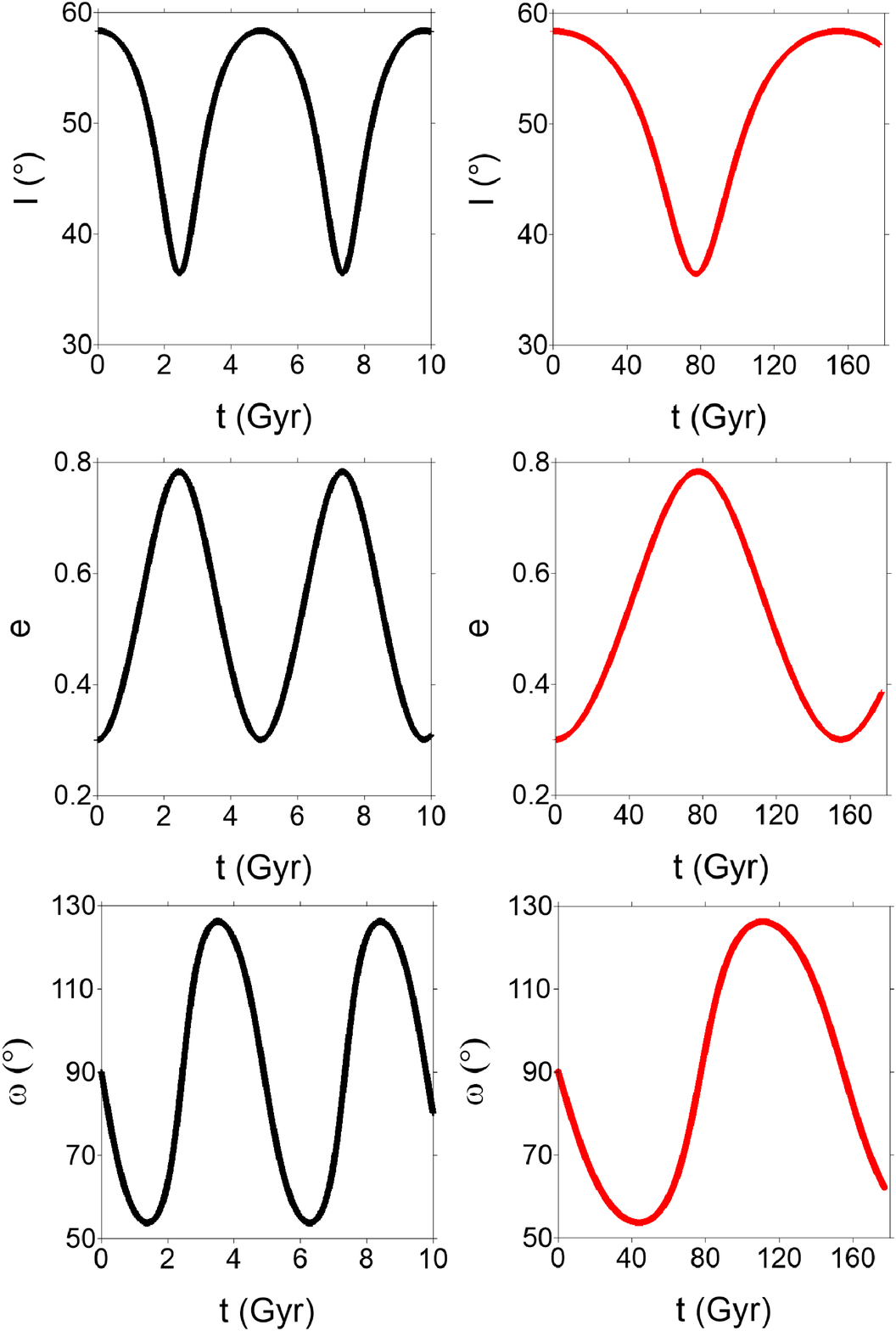}
\caption{{\small Temporal evolution of the orbital elements $I$ (top panels), $e$ (middle panels) and $\omega$ (bottom panels) of the \textit{standard system}: $a =$ 10000 au (left panels) and $a =$ 1000 au (right panels).}}
\label{fig6b}
\end{figure}

In order to test our theoretical predictions, we consider some numerical examples. From Sect. \ref{standard}, we know that the temporal evolution of the osculating ($G$, $\omega$) and mean-mean ($G^*$, $\omega^*$) action-angle variable are similar. Then, in Figures \ref{fig6a} and \ref{fig6b} we show the dynamical behaviour of first, the standard system (10000 au) and, second,  a modify standard system (1000 au). The structures of both systems have similar dynamical behaviours, which can be seen in the plane $(\omega,G/L)$  in Figure \ref{fig6a}. However, the temporal evolution of $\omega$ and $G/L$ in Figure \ref{fig6b} show different secular periods for each one. For clarity, it is easier to work with the eccentricity than with $G/L$, so in Figure \ref{fig6b} we represent the evolution of $e$ (middle panels) and to give a more complete picture we also include the evolution of $I$ (top panels). We can see that the standard system completes the cycle in 5 Gyr (top panel), but the modify standard system (bottom panel) has a period of 160 Gyr, with a relation of $160/5 \sim 32$, which confirms our prediction. The implication of this result is that a binary system with $a<1000$ au have a dynamical behaviour similar to that shown by a wide binary ($a>1000$ au), but the period of the first system has a Galactic tidal cycle longer than the age of the universe, which is in agreement with our theoretical predictions, and cannot be detected in real systems.

\begin{figure}
\centering
\includegraphics [width=0.45\textwidth] {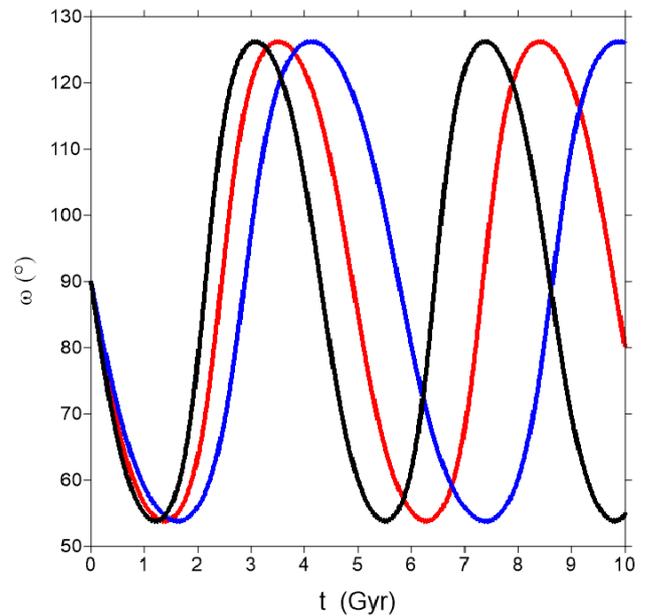}
\caption{{\small Comparison of the temporal evolution of $\omega$ for a \textit{standard system} with $m_2$:  0.8 ($A$, green line), 0.3 ($B$, red line), and 0.001 ($C$, black line) M$_\odot$.}}
\label{fig7}
\end{figure}

The other variables (or parameters) in $L^*$ are the masses of each body, $m_1$ and $m_2$. The dependence of the secular frequency with these parameters is weaker than the dependence with the semimajor axis. This is because the range of possible masses is smaller than the range in $a$ (i.e. between 10 and 0 M$_\odot$), and the dependence of the secular period with the semimajor axis has a steeper slope than with the masses. As an example, we consider the \textit{standard system} of Section \ref{standard} with a central  body of 1 M$_\odot$ and semimajor axis 10000 au, but we change the mass of the secondary: 0.8 M$_\odot$ ($A$), 0.3 M$_\odot$ ($B$), and 0.001 ($C$) M$_\odot$. The theoretical prediction of the ratio between their tidal secular periods in ${\dot{\omega^*}}$ is

\begin{equation}
\begin{array}{ccccccl}
   \\
\dfrac{\dot{\omega^*_B}}{\dot{\omega^*_A}} &=& \dfrac{T_A}{T_B}  &\sim& \sqrt{\mu_A/\mu_B} &\sim& 1.18 \,  \rm{,}\\
\\
\dfrac{\dot{\omega^*_B}}{\dot{\omega^*_C}} &=& \dfrac{T_C}{T_B}  &\sim& \sqrt{\mu_B/\mu_C} &\sim& 1.14 \,  \rm{,}\\
\\
\dfrac{\dot{\omega^*_C}}{\dot{\omega^*_A}} &=& \dfrac{T_A}{T_C}  &\sim& \sqrt{\mu_C/\mu_A} &\sim& 0.74 \,  \rm{,}\\ \,  
 \label{eq36}
\end{array}
\end{equation}

\noindent respectively. The same difference is obtained for the frequency of $G^*/L^*$. So, the difference is small and the long tidal periods for different binary systems such as star-star, star-planet, or even Sun-comet is similar. Figure  \ref{fig7} shows the temporal evolution of the osculating angle $\omega$ for the three examples, where, once again, we consider the temporal evolution of the osculating elements similar to the mean-mean action-angle variable (Sect. \ref{standard}).  The numerical determined periods are $A$: $\sim 6.1$ Gyr, $B$: $\sim $ 5.2 Gyr and $C$: $\sim $ 4.5 Gyr, with a ratio between these periods that confirms our theoretical prediction.

Then, the total mass of the binary system (i.e. $m_1$+$m_2$) is not important and has a negligible influence in the dynamical evolution of the system in the Galaxy. This is an important result because it shows that the one-degree of freedom model can be applied to different binary systems (i.e. star-star, planet-star, Sun-comet, etc.) with similar results.

 \begin{figure}
\centering
\includegraphics [width=0.45\textwidth] {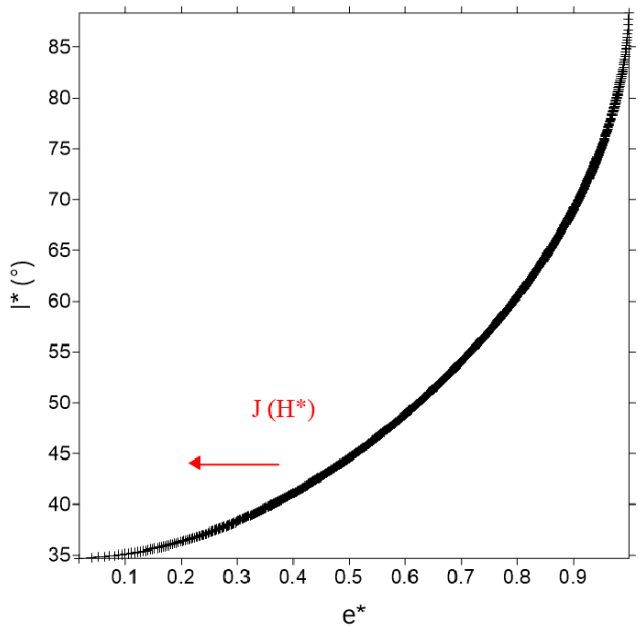}
\caption{ {\small Position of the maxima for all the possible values of $J$ (FES) for the $L^*$ of the \textit{standard system} in the plane ($e^*, i^*$). We can see that $J$ ($H^*$) increases with the decrease of $e^*$ and $I^*$. }}
\label{fig8}
\end{figure}

Finally, from our study we can deduce that the DPAM,  $J$, defines the structure of the phase space and the scale parameter $L^*$ regulates the temporal evolution of the binary system in the Galaxy. So, for each value of $J$ there is a maximum in the Hamiltonian, and from eq. (\ref{eq29}) we can obtain the position of such maximum in the plane ($e^*$, $I^*$).  In this way, different values of $J$ define a curve of equilibrium points in the plane ($e^*$, $I^*$), which can be defined as a "family of equilibrium solutions" (FES). Figure \ref{fig8} shows the position of all the equilibrium solutions calculated for the scale parameter $L^*$ of the \textit{standard system} of Section \ref{standard}. We can see that  $J$ increases when $e^*$ and $I^*$ decrease. For simplicity we plotted only the positive values of $J$ because owing to the potential symmetry the curve is similar for $I<$ 0$^\circ$. The structure of the FES in the plane ($e^*,i^*$) is equal for any value of $L^*$ because it has no influence in the dynamic behaviour of the binary systems.

\section{Application to the problem of stellar collisions}\label{referi}

 \begin{figure}
\centering
\includegraphics [width=0.45\textwidth] {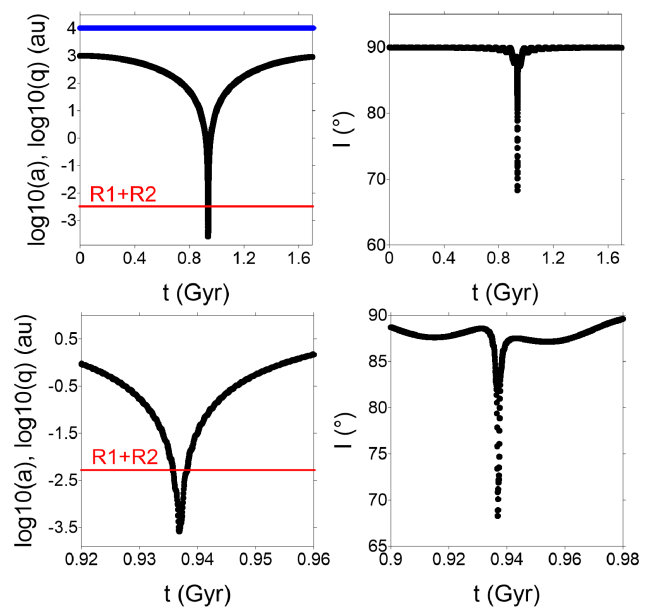}
\caption{{\small Temporal evolution of a binary system with small pericentre distance, similar to that of Fig. 1 in (Kaib \& Raymond 2014). In the top left panel we plot the semimajor axis (blue line) and the pericentre distance (black line). The red line indicates the limit of stellar collision. The top right panel shows the inclination. Bottom panels show a zoom over the minimum distance of approach between the stars. }}
\label{fig17b}
\end{figure}

The collision of the components of a wide binary star system due to the perturbation of the Galaxy (Kaib \& Raymond 2014) is an interesting  problem to which we can apply the predictions of our model. The tidal field of the Milky Way influences the pericentre distance ($q$) of the system, forcing close passages between the components of the pair.

For a collision a close encounter between the components of the binary system at a  distance less than the combined radii of the stars is necessary. The range of radii considered in (Kaib \& Raymond 2014) is $\in$ (0.13, 4.4) R$_\odot$. Then, the minimum distance of pericentre is equal to the combined radius of the stars as follows: $q_m = R_1+R_2$. According to the range of radii, the distance $q_m$ has a maximum of $\sim$ 8.8 R$_\odot$ for two massive stars, and a minimum of $\sim$ 0.26 R$_\odot$ for late-type stars. Moreover, the separation between the stars in (Kaib \& Raymond 2014) has a range of values $\in$ (1000, 30000) au. These two distances (i.e. $a$ and $q_m$) define the parameter $1-e=q_m/a$, which regulates the probability of a collision. 

We can define two limits of probability for the collision: an upper limit of maximum probability for massive stars with $a=$ 1000 au and a lower probability limit for late-type stars separated by 30000 au. These limits give us a critical value for the  eccentricity of $1-e_{c1} \sim  4 \times 10^{-5}$ in the first case and $1-e_{c2} \sim 4 \times 10^{-8}$ in the second case. 

 \begin{figure}
\centering
\includegraphics [width=0.45\textwidth] {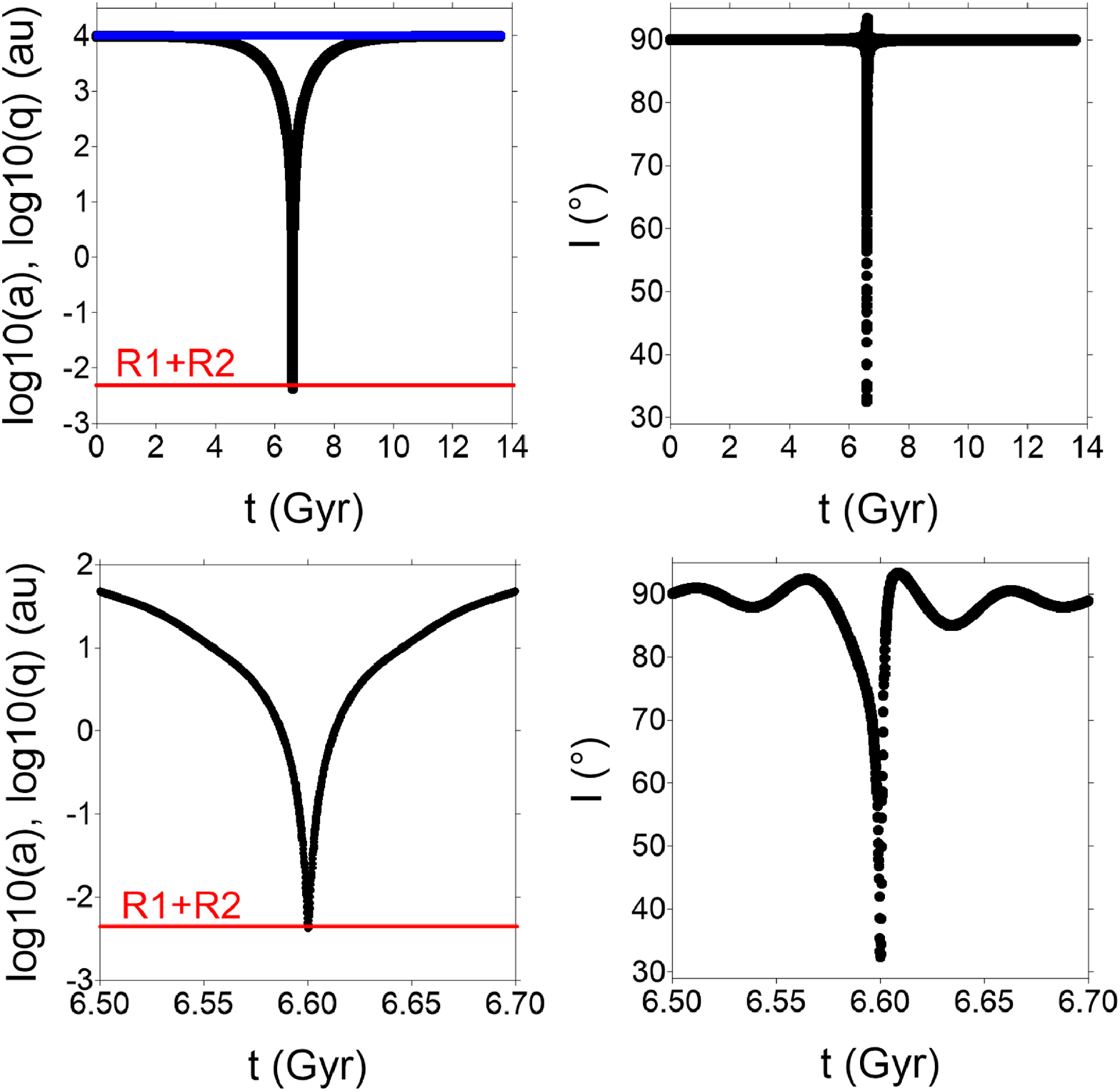}
\caption{{\small Same as Fig. \ref{fig17b}, but for a system with an initially circular orbit.}}
\label{fig17a}
\end{figure}

According to our one-degree of freedom model, the parameter that regulates the secular phase space of any binary system in the Galaxy is $J$. This parameter is defined as $J=\sqrt{1-e^2} \cos I$, which together with $e_c$ and the limit of the cosine function (i.e. $\mid \cos I \mid \, < 1$) define the probability of collision of a particular binary star system.

In both cases, $e_{c1}$ and $e_{c2}$, the maximum value of the relation $\mid J \mid/ \, \sqrt{1-e_c^2} $ is 1, which corresponds to $I=$ 0$^\circ$. This give us the critical values $\mid J_{c1} \mid \, \sim 0.01$ and  $\mid J_{c2} \mid \, \sim 3 \times 10^{-4}$.

Because the whole secular phase space is in the range $\mid J \mid \, \in (0,1)$, the limits $\mid J_{c1} \mid \,$ and $\mid J_{c2} \mid \,$ indicate the fraction of the phase space where a collision is possible. Then, we can consider $\mid J_{c1} \mid \,$ and $\mid J_{c2} \mid \,$ as the maximum and minimum probabilities of collision (i.e. $P_M$ and $P_m$) of each system. Because we are considered two limit cases, the probability of collision for any other binary stars systems is between  $ \mid J_{c1} \mid \,$ and $\mid J_{c2} \mid \,$ (i.e. between $P_M$ and $P_m$). 

We note that for a particular value of $J,$ there is a secular plane of motion associated; see Fig. \ref{fig4a} and \ref{fig4b}. Then, for particular values $\mid J_{c1} \mid$ and $\mid J_{c2} \mid$, not all the orbits in the plane achieve the high eccentricities necessary to collide, and it is very difficult to estimate the percentage of close encounters in each plane. Thus, for the sake of simplicity we give an approximate 50 \% probability of each plane. This estimation reduces the maximum $P_M$ and minimum $P_m$ probabilities by a factor two. Hence, we can conclude that a particular wide binary star system has a probability between 1 in 200 and 1 in 6000 of having a close encounter that ends in a collision between their components. Finally, although we have not considered stellar passages and tidal forces between the stars, our analytical results are in agreement with the empirical collision probability found by  \cite{kaib14}.

As an application example of the prediction of our model, we carry out two numerical simulations by integrating the exact equations of motion through a Bulirsch-Stoer code with an adopted accuracy of $10^{-13}$. For example, we consider the binary system of Figure 1 of (Kaib \& Raymond 2014) with two stars of 1 M$_\odot$ each, where $a=$ 10000 au and $q_0=$1000 au (i.e. initial pericentre distance). From the prediction of our one-degree of freedom model, this system has a collision probability of $\sim$ 0.001, which indicates that $J \leq 0.001$. Then, we arbitrarily choose $J = 8 \times 10^{-4}$, which, combined with the initial eccentricity (0.9), allow us to calculate the initial inclination, $I_0 \sim$ 89.7 $^\circ$. Our model also predicts that the angles $M$ and $\Omega$ can be chosen randomly because they are fast angles related to the secular period. We chose an initial value of 0 for the angle $\omega$ because at 90$^\circ$ it achieves the minimum pericentre distance. However we can randomly choose any value between 0 and 80, close to the minimum $q$. It is worth noting the simplicity of the process to define the required initial conditions through the one-degree of freedom model.

Moreover, we consider another example with same masses and semimajor axis, but an initially circular orbit (i.e. $q_0=$10000 au). The aim of this second example is to show the importance of the Galaxy even in circular orbits, according to the prediction of our model. The $J_c$ is equal, so we consider the same value of $J$ used to calculate the inclination. The angle $\omega$ is zero and the other two angles are randomly defined.

Figures \ref{fig17b} and \ref{fig17a} show the temporal evolution of the semimajor axis, the pericentre distance and the inclination for both examples. The red line shows the distance of collision between the stars. Although we do not consider stellar passages and the dynamical tidal force between the stars, the isolated effect of the Galactic potential allows us to deduce important conclusions:

\begin{itemize}
\item The fast angles are not important for the problem and, although we show one example for each simulation here, we have tested other values with identical results.
\item The two systems spend most of the time separated by most of 100 au and at high inclination ($I>$ 88$^\circ$). 
\item The time of approach between the stars is short compared with the total secular cycle and the inclination only decreases in such instances. 
\end{itemize}
These conclusions seem to indicate that the most important condition for a close approach between the stars are the high inclination between the orbital plane and the mid-plane of the Galaxy.

From these conclusions, we can say that the secular model  is an important tool to obtain fast results in wide binary systems. Through our model we are able to estimate the percentage of the system that ends in collision and also we easily found initial conditions that end in a collision between the stars.

\section{Conclusions}\label{conclu}

In this paper, we found an analytic one-degree of freedom model  for the secular evolution of a binary system in the Galactic environment of the solar neighbourhood. With  this model, we develop a detailed dynamic portrait of the phase space of a binary system affected by the tidal field of the Galaxy.

The simple secular model was obtained through an averaging  process applied to the three-dimensional Galactic potential of Binney \& Tremaine (2008). The mean-mean Hamiltonian and corresponding equations of motion allow a fast calculation of the secular evolution of binary systems with a low computational cost, despite a fast characterization of the phase space. We note that our reduced model is useful not only for binary stars systems, but also for  restricted two-body problems (i.e., star-comet, star-planet, etc).

On the other hand, we found that the structure of the phase space of binary systems disturbed by the tidal field of the Galaxy is only dependent on the angular momentum and is independent of the masses and the separation of the pair. Moreover, the quasi-conservation of the z component of the angular momentum gives a dynamical structure similar to the Lidov-Kozai resonance, but without the presence of a separatrix (i.e. there is not a resonant motion). This result has two important consequences for studies of binary systems: first, the Galaxy can excite the astrocentric orbit of the companion from a circular initial orbit to a high eccentric orbit (Sect. \ref{referi}), and, second, in the limit of application of this simple model we do not find a source of chaos in the secular behaviour (i.e. there is not separatrix), therefore the disruption of the pair seems to be unlikely.

However, the secular periods change as a function of the semimajor axis ($a$). Then, binary systems with lower separation between their components have higher secular periods than more separated pairs. This result is very important because it represents an analytical  dynamical interpretation for the empirical limit of $\sim$ 1000 au that separates "wide" and "tight" binary systems. For object couples with separations lower than 1000 au, the secular tidal period is greater than the age of the universe. 

We also apply our model for the study of a specific problem: stellar collision in wide binary stars systems. We have found that the secular model allows us to estimate quickly the region of the phase space where a close approach is possible. Moreover, these results may have an important application for studies of planetary systems in wide binary stars.

Finally, we have determined a limit of the application of our model, which depends on the masses and separation between the components of the binary. For very separated binary systems with small masses, the model loses precision. So, the simplifications are no longer valid and we must consider the three-dimensional complete model.

\section{bibliography}

Antognini, J. M. O. 2015, MNRAS, 452, 3610A

Antognini, J. M. O. \& Thompson, T. A. 2016, MNRAS, 456, 4219A

Bahcall, J. N., Hut, P., \& Tremaine, S. 1985, AJ, 290, 15

Binney, J. \& Tremaine, S. 2008, Galactic Dynamics: Second Edition (Princeton University Press)

Brouwer, D. \& Clemence, G. M. 1961, Methods of celestial mechanics (New York Academic Press)

Brunini, A. 1995, A\&A, 293, 935

Eggers, S. \& Woolfson, M. 1996, MNRAS, 282, 13

Ferraz-Mello, S. 2007, Canonical Perturbation Theories: Degenerate Systems
and Resonance (Springer Press)

Fouchard, M., Froeschlé, C., Valsecchi, G., \& Rickman, H. 2006, CeMDA, 95,
299

Heggie, D. C. 1975, MNRAS, 173, 729

Heggie, D. C. 2001, in The Restless Universe, ed. B. A. Steves \& A. J. Maciejew-
ski, 109–128

Heisler, J. \& Tremaine, S. 1986, ICARUS, 65, 13

Jiang, Y. F. \& Tremaine, S. 2010, MNRAS, 401, 977

Kaib, N. A. \& Raymond, S. N. 2014, ApJ, 782, 60

Kaib, N. A., Raymond, S. N., \& Duncan, M. 2013, Nature, 493, 381

Kaib, N. A., Roskar, R., \& Quinn, T. 2011, Icarus, 215, 491

Kozai, Y. 1962, AJ, 67, 591

Levison, H. F. \& Dones, L. 2001, AJ, 121, 2253

Lidov, M. L. 1961, Iskusst. sputniky Zemly 8, Acad. of Sci., U.S.S.R

Michtchenko, T. A., Lazzaro, D., Ferraz-Mello, S., \& Roig, F. 2002, Icarus, 158,
343

Morbidelli, A. 2002, Modern Celestial Mechanics: aspects of solar system dy-
namics (CRC Press)

Murray, C. D. \& Dermott, S. F. 1999, Solar system dynamics (Canbridge Uni-
versity Press)

Naoz, S. 2016, ARA\&A, 54, 441N

Naoz, S., Farr, W. M., Lithwick, Y., Rasio, F. A., \& Teyssandier, J. 2013, MN-
RAS, 431, 2155

Roell, T., Neuhäuser, R., Seifahrt, A., \& Mugrauer, M. 2012, A\&A, 542, A92

\section*{Acknowledgements}


 The authors are grateful to an anonymous referee for numerous suggestions and corrections that have helped to improve this paper.




\bibliographystyle{aa}
\small

\end{document}